\documentclass[%
 reprint,
 twocolumn,
 amsmath,amssymb,
 aps,
prb,
]{revtex4-2}

\usepackage[dvipdfmx]{graphicx}% Include figure files
\usepackage{dcolumn}% Align table columns on decimal point
\usepackage{bm}% bold math
\usepackage{color}
\usepackage{braket}
\usepackage{lipsum}
\usepackage{mathdots}
\usepackage[version=3]{mhchem} 
\usepackage{ulem}
\usepackage{color, soul}
\usepackage{xspace}
\usepackage{url}
\usepackage{appendix}
\newcommand{\red}{\textcolor{red}}

\newcommand{\green}{\textcolor{green}}

\preprint{APS/123-QED}

\begin{document}

\title{Topological spin...}% Force line breaks with \\

\date{\today}% It is always \today, today,

\title{
Nonlinear spin-motive force driven by mixed-space quantum geometry
}

\author{Tomonari Meguro$^1$}
 \email{meguro.tomonari@mbp.phys.kyushu-u.ac.jp} 
\author{Hiroaki Ishizuka$^2$}%\thanks{nomura.kentaro@phys.kyushu-u.ac.jp}}
 \email{ishizuka@hiroishizuka.com}
\author{Kentaro Nomura$^{1,}$$^3$}%\thanks{nomura.kentaro@phys.kyushu-u.ac.jp}}
 \email{nomura.kentaro@mbp.phys.kyushu-u.ac.jp}

 \affiliation{%
$^1$Department of Physics, Kyushu University, Fukuoka 819-0395, Japan \\
$^2$Department of Physics, Institute of Science Tokyo, Meguro, Tokyo 152-8551, Japan \\
$^3$Quantum and Spacetime Research Institute, Kyushu University, Fukuoka, 819-0395, Japan
}%

\begin{abstract}
Spin-motive force, i.e., the electric current induced by magnetization dynamics, is theoretically studied beyond the Thouless-pump paradigm.
In contrast to the linear-response regime, where the induced current is purely AC, we show that spin-motive force acquires both a DC component and a second-harmonic component at nonlinear order in magnetization dynamics. 
We further clarify that both contributions originate from the geometric properties of electronic bands ---  quantum geometry defined in the mixed parameter space $({\bm k}, {\bm m})$ spanned by electron's momentum ${\bm k}$ and magnetization ${\bm m}$.
By applying the theory to a Luttinger model, we demonstrate that our mechanism yields a finite nonlinear current even in the insulating regime, and the resulting electrical signal is measurable in a conventional current-measurement setup.
Our findings offer a new operating principle of AC-to-DC conversion with magnetic materials, highlighting the pivotal role of the $({\bm k}, {\bm m})$-mixed space quantum geometry in magnetization-dynamics-induced electric currents.
\end{abstract}

 \maketitle

\section{Introduction}\label{sec:intro}

The efficient and flexible interconversion between spin and charge degrees of freedom is one of the central aims of spintronics~\cite{Sinova2015,Saitoh2017, Manchon2019}.
Spin-to-charge conversion, in particular, provides an electrical route to probe and utilize the magnetization dynamics in magnetic materials and junction systems.
A paradigmatic example is the spin-motive force (SMF), in which time-dependent magnetization can generate an electromotive force and drive an electric current.
Originally, the SMF was discussed for spatially inhomogeneous magnetic textures, such as domain walls and skyrmions~\cite{Barnes2007,Duine2008,Tserkovnyak2008,SAYang2009,Shibata2011,RCheng2012,Ohe2013,Shimada2015,KMDHals2015,Yamane2019skyrmion}, and has since been experimentally observed in a wide range of materials~\cite{SAYang2009,Hai2009, Yamane2011, Schulz2012, Hayashi2012, Tanabe2012}.
Later on, it has also been shown that a spatial gradient of magnetization is not a prerequisite; in the presence of spin-orbit interaction, a uniform magnetization undergoing temporal variation can induce an electric current~\cite{Kim2012, Tatara2013socSMF, Yamane2013socSMF, Freimuth2015}; 
they are also referred to as the SMF in a narrower sense.
Despite extensive studies of SMF in various systems, the discussion has been largely limited to the linear-response regime, where the resulting electric current is proportional to the time derivative of magnetization.
Beyond linear response, one may expect spin-to-charge conversion to enable rectification and higher-harmonic generation.
Such nonlinear effects provide key functionalities for flexible electrical readout and frequency conversion driven by magnetization dynamics—an aspect that has remained largely unexplored.

In the study of nonlinear responses, it has been increasingly recognized that the geometrical properties of electronic states, collectively termed quantum geometry, play crucial roles \cite{Xiao2010,Nagaosa2010, Bandyopadhyay2024,Liu2025, AGaoNN2025}. 
% \blue{which are often called intrinsic mechanisms}.
Key quantities that characterize the geometry of quantum states are the metric and curvature in the parameter space on which the quantum states are defined, known as the quantum metric and the Berry curvature, respectively.
%Mathematically, quantum geometry is characterized by the metric and curvature structure in the parameter space on which the electrons depend. 
Historically, the importance of the Berry curvature has long been recognized in the context of linear electrical transport \cite{Xiao2010,Nagaosa2010}, such as in the quantum and anomalous Hall effects; they are related to the Berry curvature of Bloch states 
% ~\red{\cite{Luttinger?,TKNN1982}} 
in the crystal momentum ($\bm k$) space.
%Examples include charge pumping under adiabatic evolution known as the Thouless pump \cite{Thouless1983, QNiu1984}, where the parameters are the time $t$ and the electron momentum $k$.
%The others, the integer and anomalous Hall effects, are exhibited by the Bloch electron's geometry \red{\cite{TKNN1982,}} whose states depend on the crystal momentum $\bm k$. 
Recently, the role of the quantum metric in nonlinear responses has also been actively discussed~\cite{Bandyopadhyay2024,Liu2025, AGaoNN2025}. 
Studies of the nonlinear Hall effect~\cite{YGao2014,Sodemann2015,CXiao2019NLHE} have revealed that the response coefficients can be expressed in terms of the quantum metric or the Berry curvature-dipole.
% 
% \red{
These findings have offered not only a new way of understanding electrical responses, but also design principles for functional materials and electronic devices with rectifying and high-frequency sensing capabilities \cite{YZhang2021, BCheng2024, YOnishi2024}.
% }
% 
To date, the role of $\bm k$-space quantum geometry in material responses has been well established, both theoretically and experimentally.
%More recently, attention has also turned to the role of the quantum metric in nonlinear regimes\cite{Bandyopadhyay2024,Liu2025}. 
%Studies of nonlinear Hall effect\cite{YGao2014,Sodemann2015,CXiao2019NLHE} have revealed that the response coefficients can be expressed in terms of the the quantum metric or the Berry curvature. 
% 
%Up to date, the $\bm k$-space quantum geometry has been well established theoretically and experimentally.
% especially in the electron's momentum space, it has been well established that various response phenomena arise from the quantum geometry.

% \red{
These developments motivate revisiting spin-charge conversion from a quantum geometric viewpoint.
Extending this viewpoint to spintronics naturally leads to the quantum geometry of the ($\bm k, \bm m$)-mixed space, in which the magnetization $\bm m$ is considered an additional parameter. %of the Bloch electron's states. 
% }
Indeed, the observation of intrinsic spin-orbit torque \cite{HKurebayashi2014} 
% in (Ga,Mn)As-based ferromagnetic semiconductors, followed by 
and a subsequent theoretical analysis~\cite{Freimuth2014} opened a pathway toward understanding charge-spin conversions as a phenomenon related to the $(\bm k, \bm m)$-mixed space quantum geometry 
\cite{Freimuth2015,Hanke2017,BXiong2018,CXiao2021,CXiao2022_2ndSOT,CXiao2023,JTang2024,Manchon2024,XFeng2025,YRen2025,Meguro2025}.
% , much like the role played by momentum-space geometry in charge transport.
% 
% 
For SMF, theoretical studies have proposed that the effect originates from the Berry curvature defined in the  $(\bm k, \bm m)$-mixed space \cite{Freimuth2015,JTang2024,Manchon2024}. 
However, most existing studies have focused on the linear-response regime, ${\bm j} \approx J_{\rm ex} \dot{\bm m} $, where $J_{\rm ex}$ is the exchange coupling constant between the electron spin and the magnetization.
% 
% \red{
In that regime, the induced electric current remains purely AC, without a DC component or frequency-converted components
% }
; it oscillates in time, %as sin$\omega t$, cos$\omega t$, 
and the average current over a cycle of magnetization precession vanishes. 
Furthermore, preceding theoretical works have concentrated on the Berry curvature contribution, leaving the role of the quantum metric unclarified.
%Hence, the broader quantum-geometric structure inherent to electronic states is not fully captured.
% To achieve and understand the flexible frequency conversion and electrical readout of magnetization dynamics in a unified manner, 
% it is desired to clarify the microscopic mechanism of the SMF in the nonlinear regime, based on the quantum geometry.
% A microscopic theory of nonlinear SMF based on the quantum geometry is therefore needed to clarify how rectification and frequency conversion can arise intrinsically from magnetization dynamics.
% \red{
A theory of nonlinear SMF based on mixed-space quantum geometry is needed to clarify how magnetization dynamics generates finite cycle-averaged and higher-harmonic responses, and how these signals are linked to the geometric properties of electronic states.
% }

In this work, we develop a theory of SMF in the nonlinear regime from the perspective of quantum geometry in the $(\bm k, \bm m)$-mixed space. 
Considering a ferromagnet in which 
% \red{
the spatially homogeneous magnetization varies in time
% }
, we formulate the electric current response to second order in the magnetization dynamics, using semiclassical wave-packet theory.
% and a perturbative expansion in the exchange interaction.
% 
Consequently, we find that a nonlinear SMF exhibits DC and second harmonic generation (SHG) under precessional magnetization dynamics. 
% 
% \red{
We further demonstrate that their geometric origins are distinguished by frequency scaling; 
contributions linear in the precession frequency are governed by the Berry curvature,
whereas those quadratic in frequency are governed by the quantum metric, 
both of which are defined in the $(\bm k, \bm m)$-mixed space.
% }
% 
To illustrate the significance of the nonlinear SMF, 
we perform numerical calculations in a Luttinger model. 
As a result, we find that, even in the insulating regime, 
% \red{
our mechanism yields a finite nonlinear SMF, and the resulting signal is detectable in a standard current measurement.
% }
% 
Our findings suggest that quantum geometry in the mixed space plays a pivotal role in the field of nonlinear spintronics.

% offers a new path way for the low dissipative spintronics. 

\section{Wave-packet theory}\label{sec:general}

\subsection{Wave-packet}

To study the effect of magnetization dynamics on electron dynamics, we consider a Hamiltonian that consists of a static term and time-dependent terms which are treated perturbatively. The Hamiltonian reads,
\begin{align}
    {\hat H}(t) = {\hat H}^{(0)} + \sum_\beta \delta m_\beta(t) {\hat H}'_\beta %+ -e\phi + \cdots
     .\label{eq:Hamil}
\end{align}
where $\beta=x,y,z$. % labels a coordinate in the magnetization space. 
We assume the static term ${\hat H}^{(0)}$ consists of temporally uniform terms, such as the electron kinetic energy, the crystal potential, and the static magnetization $\bm m_0$, and potentials that vary slowly compared to the lattice spacing, 
%other spatial modulation fields 
such as the electromagnetic potential by external fields.
% which only depends on a spatial coordinate.
% 
%The length scale on which the modulation fields vary is assumed to be much larger than the length of the lattice constant.
% 
In the time-dependent parts, $\delta m_{\beta}(t)$ is assumed to be small, $\delta m\ll m_0$ ($\delta m=|\delta{\bm m}(t)|$ and $m_0=|\bm m_0|$).
In addition, 
$\delta m_{\beta}(t)$ is periodic in time with $T$ and changes slowly such that $\hbar/T$ is sufficiently small compared to the energy scale of $\hat H^{(0)}$.

Following the semiclassical theory for electron dynamics \cite{Sundaram1999,YGao2014}, 
we consider the dynamics of a wave-packet-type wavefunction localized at $\bm x_c, \bm k_c$ in the Liouville space. %at a given time ; $\bm x_c(t), \bm k_c(t)$. 
Here, $\bm x_c$ and $\bm k_c$ are the center of mass of the wave packet in the real and momentum space, respectively.
The time-evolution of the wave packet is determined by a variational method. % which \green{we} will \green{describe in the following}.
To this end, we consider a situation such that %the $\bm x_c$-dependence enters into ${\hat H}^{(0)}$ as electrostatic potential $-e\phi({\bm x}_c)$
the slowly-varying potential in ${\hat H}^{(0)}$ is the electrostatic potential $\phi({\bm x})$
\footnote{
One can consider the Hamiltonian has $\bm x_c$-dependencies other than the electrostatic potential, such as vector potential and lattice distortion.
In this situation, 
we define the local Hamiltonian and a gradient one by expanding the Hamiltonian around $\bm x_c$ ; 
${\hat H}^{(0)} = {\hat H}^{(0)}_c + \Delta{\hat H}^{(0)}$ 
where ${\hat H}^{(0)}_c$ only contains $\bm x_c$, and $\Delta{\hat H}^{(0)}$ contains the power of the spatial gradient $\partial_{{\bm x}_c}$.
Because of the length scale of the modulation field mentioned above, 
${\hat H}^{(0)}_c$ is lattice periodic locally,
and therefore we can obtain its Bloch eigenstates and values at each point $\bm x_c$ locally.
}.
Defining the Bloch Hamiltonian for ${\hat H}^{(0)}$ as ${\hat {\mathcal H}}^{(0)}_{{\bm k}} = e^{i{\bm k}\cdot{\hat{\bm x}}} {\hat H}^{(0)} e^{-i{\bm k}\cdot{\hat{\bm x}}}$, one can obtain the Bloch eigenstates and values as 
\begin{align}
    {\hat {\mathcal H}}^{(0)}_{{\bm k}} \ket{ u^{(0)}_{n{\bm k}}({\bm m}_0)} 
      &= 
    \varepsilon^{(0)}_{n{\bm k}}({\bm x}_c, {\bm m}_0) \ket{ u^{(0)}_{n{\bm k}}( {\bm m}_0)}, 
\end{align}
where $\bm k$ is the crystal momentum. 
The electrostatic potential merely shifts the eigenvalues, and the Bloch eigenstates do not depend on $\bm x_c$.
% 
% 
% 
% 
% In the semiclassical theory for electron dynamics, we consider the dynamics of a wave-packet-type wavefunction, which is localized both in the real and momentum spaces,
% 
Using this local Bloch eigenstates,
we construct the wave-packet by %by using the local Bloch eigenstates obtained above as a basis system, 
\begin{align}\label{eq:WavePacket}
     \ket{ \Psi_{{\bm x}_c, {\bm k}_c}} &= \int_{\bm k} e^{i {\bm k}\cdot{\hat{\bm x}}}
                               \Bigl[~
                                     \sum_{l} 
                                     C_{l{\bm k}}({\bm x}_c, {\bm k}_c,t)\,
                                     \ket{ u^{(0)}_{l{\bm k}}({\bm m}_0)}~
                                       \Bigr],
\end{align}
% 
% Here, ...
% \textcolor{blue}{
% (the momentum integral), (the explanation of the coefficients), (Should we separate the intraband/interband contributions explicitly?).
% }.
where $\int_{\bm k} \equiv \int_{\rm BZ} d^d{\bm k}/(2\pi)^d$ denotes the momentum integral in the Brillouin zone (BZ).
The coefficients $C_{l{\bm k}}({\bm x}_c, {\bm k}_c,t)$ are chosen such that the wave function is 
sharply peaked at $\bm x_c$ in the real space and at $\bm k_c$ in the momentum space,
%The wave-packet-type wave function must yield the preassigned centers of mass as  
i.e., ${\bm x}_c = \bra{\overline{\Psi_{{\bm x}_c, {\bm k}_c}}} \hat {\bm x} \ket{\overline{\Psi_{{\bm x}_c, {\bm k}_c}}}$ and 
${\bm k}_c = \bra{\overline{\Psi_{{\bm x}_c, {\bm k}_c}}} \hat {\bm k} \ket{\overline{\Psi_{{\bm x}_c, {\bm k}_c}}}$, 
where 
\begin{align}
\ket{\overline{ \Psi_{{\bm x}_c, {\bm k}_c}}}
     &= \frac{\ket{{ \Psi_{{\bm x}_c, {\bm k}_c}}}}
            {\sqrt{\langle \Psi_{{\bm x}_c, {\bm k}_c} | \Psi_{{\bm x}_c, {\bm k}_c} \rangle}}
\end{align}
%$\ket{\overline{\Psi_{{\bm x}_c, {\bm k}_c}}}$
is the normalized wave-packet.
In most cases, the wave-packet is assumed to be confined in a band, which we denote as the $n$-th band, i.e., $C_{l{\bm k}}({\bm x}_c, {\bm k}_c,t)=0$ ($l\ne n$).
In this work, however, we consider the effect of inter-band transition, which we assume to be $|C_{l{\bm k}}({\bm x}_c, {\bm k}_c,t)|\ll |C_{n{\bm k}}({\bm x}_c, {\bm k}_c,t)|$ ($l\ne n$).

% In particular, for the momentum-center, we practically replace  
% $|C_{n{\bm k}}(\xi_{c}(t))|^2 \approx \delta({\bm k}-{\bm k}_c(t))$ by the spirits that 
% the width of the wave-packet is sufficiently narrow compared to that of the Brillouin zone.

In principle, one may choose any form for the coefficient $C_{l{\bm k}}({\bm x}_c, {\bm k}_c,t)$ for $l\ne n$ in Eq. (\ref{eq:WavePacket}).
In the current case, as a natural choice of $C_{l{\bm k}}({\bm x}_c, {\bm k}_c,t)$, 
we choose the coefficient based on the first order perturbation theory.
Therefore, we require that the wave-packet given in Eq. (\ref{eq:WavePacket}) follows the time-dependent Schr\"{o}dinger equation, 
\begin{align}\label{eq:TimeDep_Sch_eq}
    i\hbar %\frac{\partial}{\partial t}
           \partial_t \ket{ \Psi_{{\bm x}_c, {\bm k}_c}} 
         = \Bigl( {\hat{H}}^{(0)} + \sum_\beta\delta m_\beta(t) {\hat H}'_\beta \Bigr) \ket{ \Psi_{{\bm x}_c, {\bm k}_c}}.
\end{align}
Here, the time dependence of the wave-packet comes from both $C_{l{\bm k}}({\bm x}_c, {\bm k}_c,t)$ explicitly and centers of mass implicitly.
Assuming that $\sum_\beta\delta m_\beta(t) {\hat H}'_\beta$ is a small correction to ${\hat H}^{(0)}$, 
we find that the coefficient reads
% the correction to the wave function reads 
(see Appendix.\ref{sec:Appendix.A} for the derivation),
\begin{align}\label{eq:M_def}
    C_{l{\bm k}}({\bm x}_c, {\bm k}_c,t) 
      &= M_{ln}({\bm x}_c, {\bm k}_c,t)~C_{n\bm k}({\bm x}_c, {\bm k}_c,t),\\
% 
    % M^{(1)}_{ln}({\bm x}_c, {\bm k}_c,t) &= \sum_{\beta}
    %      \frac{ \varepsilon^{(0)}_{ln,{\bm k}}({\bm x}_c,{\bm m}_{0}) \,\delta m_{\beta}(t) - i\hbar \,\delta {\dot m}_{\beta}(t) }
    %           { (\varepsilon^{(0)}_{ln,{\bm k}}({\bm x}_c,{\bm m}_{0}))^2 - \hbar^{2} \omega^{2}}  
    %           ( {\hat H}'_{\beta} )_{ln}
    %      % \,\sigma^{ln}_{\beta}({\bm x}_c), 
    % 
    M^{(1)}_{ln,\bm k}(\bm x_c, \bm k_c, t) 
      &= - \sum_{\beta,\Omega} \frac{ \delta m_{\Omega,\beta}~e^{-i\Omega t} }{ \varepsilon^{(0)}_{ln,\bm k}(\bm x_c,\bm m_0) - \hbar\Omega }( {\hat H}'_{\beta} )_{ln}
\end{align}
% \red{(above formula is valid in the situation where magnetization precesses. Should we show the formula for general time-varying magnetization ?)}
% 
where $\delta m_\beta (t)$ is expanded with its discrete Fourier components, 
$\delta m_\beta(t) = \sum_{\Omega}\delta  m_{\Omega,\beta} e^{-i\Omega t}$.
We assume that $C_{l{\bm k}}({\bm x}_c, {\bm k}_c,t)$ differs from $C_{n{\bm k}}({\bm x}_c, {\bm k}_c,t)$ only by an overall factor $M_{ln,\bm k}({\bm x}_c, {\bm k}_c,t)$; this assumption is supported by the perturbation theory, in which $M^{(1)}$ gives the first order correction.
% 
%For the factor $M_{ln}$, we expand $M_{ln}({\bm x}_c, {\bm k}_c,t)$ order by order, 
We further expand $M_{ln}$ in power of $\delta m$,
$M_{ln}({\bm x}_c, {\bm k}_c,t) = \sum_{N \ge 1} M^{(N)}_{ln}({\bm x}_c, {\bm k}_c,t)$, where $M^{(N)}_{ln}({\bm x}_c, {\bm k}_c,t)$ is $\mathcal O (\delta m^{N})$.
% 
% $\beta$ labels a coordinate in the magnetization space, 
% specifically running over the transverse components of the magnetization, e.g., $\beta = y,z$ with precession around the $x$-axis.
% 
In the denominator, $\varepsilon^{(0)}_{ln,{\bm k}}({\bm x}_c,{\bm m}_{0}) = \varepsilon^{(0)}_{l{\bm k}}({\bm x}_c,{\bm m}_{0}) - \varepsilon^{(0)}_{n{\bm k}}({\bm x}_c,{\bm m}_{0})$ 
denotes the difference of energy between the eigenstates in $l$-th and $n$-th bands, and
% The superscript of $M^{(1)}_{ln}({\bm k},t)$ represents its order with respect to perturbation, 
% $M^{(N)}_{ln}({\bm k},t)\propto \mathcal O (\delta m^{N})$.
$( {\hat H}'_{\beta} )_{ln}
=\langle u^{(0)}_{l{\bm k}}({\bm m}_{0})|{\hat H}'_{\beta}|u^{(0)}_{n{\bm k}}({\bm m}_{0})\rangle$ is 
a matrix element for the perturbative Hamiltonian
% Pauli operators 
that is expressed by the unperturbed Bloch eigenstates. 

Consequently, the wave-packet wavefunction in Eq.~(\ref{eq:WavePacket}) reads 
\begin{align}\label{eq:WavePac_pertub}
     \ket{ \Psi_{{\bm x}_c, {\bm k}_c}} 
              &= \int_{\bm k} e^{i {\bm k}\cdot{\hat{\bm x}}}\,
                                       C_{n}({\bm x}_c, {\bm k}_c,t)\,
                                       \ket{ \tilde u_{n{\bm k}}({\bm x}_c,{\bm m})}, 
\end{align}
where $\ket{ \tilde u_{n{\bm k}}({\bm x}_c,\bm m) }$ are the Bloch eigenstates of the Hamiltonian ${\hat H}(t) = {\hat H}^{(0)} + \sum_\beta\delta m_{\beta}(t){\hat H}'_{\beta}$ up to the first order in $\delta m_\beta$, %of the time-dependent perturbation, 
and are given by 
$\ket{ \tilde u_{n{\bm k}}({\bm x}_c, \bm m) }
       = \ket{ u^{(0)}_{n{\bm k}} (\bm m_{0}) }
          + \ket{ u^{(1)}_{n{\bm k}} ({\bm x}_c, \bm m) } + \cdots$ 
with 
\begin{align}\label{eq:Bloch_pertub}
    \ket{ u^{(1)}_{n{\bm k}}({\bm x}_c, \bm m) } &= \sum_{l (\ne n)} M^{(1)}_{ln}({\bm x}_c, {\bm k}_c,t)\, \ket{u^{(0)}_{l{\bm k}}( \bm m_{0}) }.
\end{align}
% 
%For the calculation, it is convenient to introduce a normalized wavefunction for Eq.~\eqref{eq:WavePac_pertub}.
Using these results, the normalized wavefunction up to the second order in $\delta m$ reads,
\begin{align}\label{eq:WavePac_nomalized}
      \ket{\overline{ \Psi_{{\bm x}_c, {\bm k}_c}}}
     %&= \frac{\ket{\overline{ \Psi_{{\bm x}_c, {\bm k}_c}}}}
     %       {\sqrt{\langle \Psi_{{\bm x}_c, {\bm k}_c} | \Psi_{{\bm x}_c, {\bm k}_c} \rangle}} \notag \\
     &= \frac{\ket{{ \Psi_{{\bm x}_c, {\bm k}_c}}}}
            {\sqrt{1 + \langle \Psi^{(1)}_{{\bm x}_c, {\bm k}_c}| \Psi^{(1)}_{{\bm x}_c, {\bm k}_c} \rangle 
            + {\mathcal O}(\delta m^{3})}} \notag \\
     &= (1+\delta)\ket{ \Psi^{(0)}_{{\bm x}_c, {\bm k}_c}}
        + \ket{ \Psi^{(1)}_{{\bm x}_c, {\bm k}_c}} 
        + \ket{ \Psi^{(2)}_{{\bm x}_c, {\bm k}_c}}
        + \cdots ,
\end{align}
where $|\Psi^{(N)}_{{\bm x}_c, {\bm k}_c} \rangle \propto {\mathcal O}(\delta m^{N})$ and 
$\delta = -\tfrac{1}{2}\langle \Psi^{(1)}_{{\bm x}_c, {\bm k}_c}|\Psi^{(1)}_{{\bm x}_c, {\bm k}_c} \rangle$.

\subsection{Equations of motion}

The dynamics of the wave-packet is determined by the extrema of action, which reads
\begin{align}\label{eq:action}
S=\int dt\,L[\bm x_c(t),\bm k_c(t)],
\end{align}
where $L[\bm x_c(t),\bm k_c(t)]$ is the Lagrangian.
Using the normalized wavefunction, the Lagrangian for Eq.~\eqref{eq:WavePac_nomalized} reads 
(the detailed derivation is presented in Appendix.\ref{sec:Appendix.B})
\begin{align}\label{eq:Lagrangian}
     L &= \bra{\overline{ \Psi_{{\bm x}_c, {\bm k}_c}}} 
             \left( i\hbar\frac{d}{dt} - {\hat H}(t) \right) 
             \ket{\overline{ \Psi_{{\bm x}_c, {\bm k}_c}}} \notag \\
       &= \hbar{\dot{\bm x}}_{c} \cdot {\bm k}_{c}
        % - \tilde{\mathcal E}^{\rm WP}_{n}- \tilde{\mathcal E}^{\rm dyn}_{n}
        - \tilde{\mathcal E}_{n}
        - \phi({\bm x}_{c}) \notag \\
       &\ \ \ \ \ \ \ \ \ 
       % \red{ - \hbar{\dot{\bm x}}_{c,n} \cdot \tilde{\bm A}^{{\bm x}_{\rm c}}_{n}}
        - \hbar{\dot{\bm k}}_{c} \cdot \tilde{\bm A}^{{\bm k}_{\rm c}}_{n}
        - \hbar{\dot{\bm m}}     \cdot \tilde{\bm A}^{\bm m}_{n},
       % &=  \langle \Psi^{(0)}_{n}(t) | 
       %        i\hbar\frac{\partial }{\partial t} | \Psi^{(0)}_{n}(t) \rangle
       %    + {\dot{\bm m}}\cdot \langle \overline{\Psi_{n}(t)} |
       %        i\hbar\frac{\partial }{\partial {\bm m}} | \overline{\Psi_{n}(t)} \rangle
       %    - \tilde{\mathcal E}_{n},
\end{align}
% \blue{
where %we introduce a Berry connection that includes the perturbative correction as
$\tilde{\bm A}^{{\bm k}_{\rm c}}_n = 
-i\langle \tilde u_{n{\bm k}_{\rm c}} (\bm x_c,{\bm m})|
\partial_{{\bm k}_c} 
| \tilde u_{n{\bm k}_{\rm c}}(\bm x_c,{\bm m}) \rangle$
is the Berry connection with the perturbative correction, and 
$\tilde{\bm A}^{{\bm m}}_n$ is defined in the same manner but with the derivatives with respect to $\bm m$. 
We ignore the Berry connection with respect to $\bm x_c$ under the approximation that neglects terms of second or higher order in the product of $\delta m$ and the spatial gradient, since it is not the main focus of this study.
The energy of the $n$-th wave-packet consists of two terms,
$\tilde{\mathcal E}_{n} = \tilde{\mathcal E}^{\rm WP}_{n} + \tilde{\mathcal E}^{\rm dyn}_{n}$, where
% 
%Here we introduce the wave-packet energy $\tilde{\mathcal E}^{\rm WP}_{n}$ 
\begin{align}\label{eq:WavePacEn_def}
     \tilde{\mathcal E}^{\rm WP}_{n} &= 
           \bra{ \overline{ \Psi_{{\bm x}_c, {\bm k}_c}} }
              \hat H(t) 
            \ket{ \overline{ \Psi_{{\bm x}_c, {\bm k}_c}} }, 
\end{align}
is the wave-packet energy
and
\begin{align}\label{eq:dynEn_def}
     \tilde{\mathcal E}^{\rm dyn}_{n} &= 
          \bra{ \Psi^{(0)}_{{\bm x}_c, {\bm k}_c} } 
          i\hbar\partial_t 
          \ket{ \Psi^{(0)}_{{\bm x}_c, {\bm k}_c} }
          - \bra{ \overline{ \Psi_{{\bm x}_c, {\bm k}_c}} }
              i\hbar \partial_t 
            \ket{ \overline{ \Psi_{{\bm x}_c, {\bm k}_c}} },
\end{align}
is the dynamical energy;
the latter comes from the dynamical part of the definition of the Lagrangian.

% the quantity which has an energy-unit and comes from the dynamical part of the definition of the Lagrangian, termed as 
% the dynamical energy %$\tilde{\mathcal E}^{\rm dyn}_{n}$

By applying the principle of least action to Eq.~\eqref{eq:action}, we obtain the following equations of motion for $\bm x_c$ and $\bm k_c$,
\begin{align}\label{eq:wavepac_k}
\hbar{\dot k}_{c,i} &= -eE_{i}, \\
%     
                    % &\ \ \ \ \ 
                    %  - \hbar\sum_{j} {\tilde \Omega}^{{\bm x}_c{\bm x}_c}_{ij}{\dot x}_{c,j} 
                    %  - \hbar\sum_{j} {\tilde \Omega}^{{\bm x}_c{\bm k}_c}_{ij}{\dot k}_{c,j} 
                    %  - \hbar\sum_{\alpha}{\tilde \Omega}^{{\bm x}_c{\bm m}}_{i\alpha}{\dot m}_{\alpha}\\
% 
     \label{eq:wavepac_x}
     {\dot x}_{c,i} &= \frac{1}{\hbar}\frac{\partial \tilde{\mathcal E}_{n}}{\partial k_{c,i}} 
%     
                     % &\ \ \ \ \ 
                     % + \sum_{j} {\tilde \Omega}^{{\bm k}_c{\bm x}_c}_{ij}{\dot x}_{c,j} 
                     + \sum_{j} {\tilde \Omega}^{{\bm k}_c{\bm k}_c}_{ij}{\dot k}_{c,j} 
                     + \sum_{\alpha} {\tilde \Omega}^{{\bm k}_c{\bm m}}_{i\alpha}{\dot m}_{\alpha}.
\end{align}
Here, $ \bm E = -\bm \nabla_{\bm x_c}\phi(\bm x_c) $ is the electric field, and 
% 
%We introduce the Berry curvatures in phase space $({\bm k}_{\rm c},{\bm m})$ as
\begin{align}\label{eq:BC_def}
     \tilde\Omega^{{\bm k}_{\rm c}{\bm m}}_{n,i \alpha} 
      &= \partial_{k_{{\rm c},i}} \tilde A^{{\bm m}}_{\alpha} - \partial_{m_{\alpha}} \tilde A^{{\bm k}_{\rm c}}_{i}, 
\end{align}
is the Berry curvature in the $({\bm k}_{\rm c},{\bm m})$ phase space.
${\tilde \Omega}^{{\bm k}_c{\bm k}_c}_{ij}$ is the Berry curvature in the momentum space and defined in the same manner.

\section{Nonlinear current generation}\label{sec:NLcurrent}

% \section{Magnetic precession}\label{sec:WavePac_perturb}
%\subsection{Set up : Magnetic precession}\label{subsec:setup}
\subsection{Dynamics under magnetic precession}\label{subsec:setup}
% \textcolor{red}{
% \begin{itemize}
%     \item magnetization dynamics is precession around z-axis, precession frequency $\omega$, amplitude $\delta m$ is small $\rightarrow$ perturbation.
%     \item about Hamiltonian
%     \item preparation of non-perturbative (Bloch) states, energy
% \end{itemize}
% }
% #####################
% ・磁性体内で，$\bm m$で特徴づけられる磁化が周波数$\omega$で歳差運動をしている状況を考える．
% ・歳差運動の軸を向いた磁化ベクトルを$\bm m_{0}$，その周りの歳差運動を$\delta {\bm m}(t)$と表記することにする．
% ・ここで，歳差の振幅は$\delta m$とし，以降はこれを微小量として扱う．
% ・以降では，z軸周りの歳差運動を考えることにする．この時，磁化ベクトル$\bm m$は次のように表すことができる．
% 
Having established the formalism for the wave-packet dynamics, we now turn to studying the electric current generated by the magnetic dynamics.
%An example of the Hamiltonian in Eq.~\eqref{eq:Hamil} is 
To this end, we consider the electrons subject to the magnetic precession of the ferromagnetic moment.
Suppose the ferromagnetic moment ${\bm m}(t)$ precess about ${\bm m}_{0}$, i.e., ${\bm m}(t)={\bm m}_{0}+\delta{\bm m}(t)$,
where $\delta {\bm m}(t)$ is assumed to be small, $\delta m\ll m_0$ ($\delta m=|\delta{\bm m}(t)|$ and $m_0=|\bm m_0|$).
%The average magnetization is denoted $\bm m_{0}$, and the deviation from $\bm m_0$ is written as $\delta {\bm m}(t)$. % ; ${\bm m}(t)={\bm m}_{0}+\delta{\bm m}(t)$.
%The amplitude $\delta m=|\delta{\bm m}(t)|$ is assumed to be small compared to $|\bm m_{0}|$. 
% 
% 
% For this magnetization configuration, the total Hamiltonian consists of a static part and a time-dependent perturbation as ${\hat H}(t) = {\hat H}_{0} + {\hat H}_{1}(t)$. 
% 
The unperturbed term is given by ${\hat H}^{(0)} = {\hat H}_{e} - J_{\rm ex}{\bm m}_{0}\cdot{\bm{\hat \sigma}}$, which is composed of the electronic Hamiltonian ${\hat H}_{e}$ and the exchange coupling aligned with the magnetization at the equilibrium state.
Here, $J_{\rm ex}$ is the coupling constant between itinerant electrons and $\bm m_0$.
The perturbation, $\sum_\beta\delta m_{\beta}(t) {\hat H}'_{\beta} = - J_{\rm ex} \sum_\beta\delta m_{\beta}(t){\hat \sigma}_{\beta}$, accounts for the exchange coupling arising from the deviation $\delta\bm m(t)$.
% transverse components of the precession.

%It is sufficient to consider precession around one of the principal axes as the axis of magnetization dynamics.
For the sake of concreteness, let us consider a magnetization precession about the $x$-axis with angular frequency $\omega$.
%As an example, we consider a magnetization precessing at angular momentum $\omega$.
% 
% While the following formulation can be extended to an arbitrary precession axis, we restrict ourselves to the $z$-axis for notational convenience.
% 
In this case, the axis vector is ${\bm m}_{0}=(\sqrt{1-(\delta m)^2},0,0)^{\rm T}$, and the transverse component takes the form
% For example, for the precession about the $z$ axis,
% ${\bm m}_{0}=(0,0,\sqrt{1-(\delta m)^2})^{\rm T}$, and 
\begin{align}\label{eq:mag_vec}
     \delta {\bm m}(t) &= \delta \bm m^{(+)}e^{-i\omega t} + \delta \bm m^{(-)}e^{i\omega t},
                         %  \frac{\delta m}{2} \left\{
                         %  \begin{pmatrix}
                         %    0 \\
                         %    1 \\
                         %    i \\
                         %  \end{pmatrix} e^{-i \omega t}
                         % +\begin{pmatrix}
                         %    0 \\
                         %    1 \\
                         %    -i \\
                         %  \end{pmatrix} e^{i \omega t} \right \}.
\end{align}
where $\delta \bm m^{(\pm)} = \frac{\delta m}{2}(0,1,\pm i)^{\rm T}$.
%It highlights the circularly polarized components of the precession. 
For later convenience, we introduce its Fourier components
% $\delta m_\beta(t) = \sum_{\Omega}\delta  m_{\Omega,\beta} e^{-i\Omega t}$
% $\delta \bm m(t) = \int \frac{d\Omega}{2\pi} e^{i\Omega t}\delta \bm m(\Omega)$  
% 
\begin{align}\label{eq:mag_frequency}
    \delta \bm m_\Omega = \delta \bm m^{(+)} \delta_{\Omega,\omega}+\delta \bm m^{(-)} \delta_{\Omega,-\omega}.
\end{align}

%\subsection{Wave packet dynamics}

Applying the wave packet theory in Sec.~\ref{sec:general}, 
under the magnetization precession, 
we find that the wave-packet dynamics reads
\begin{align}
     {\dot k}_{{\rm c},i} &= 0, \\
     \label{eq:wavepac_x_EOM}
     {\dot x}_{{\rm c},i} &= \frac{1}{\hbar} \frac{\partial \tilde{\mathcal E}_{n}}{\partial k_{{\rm c},i}} 
                     + \sum_{\alpha}\tilde\Omega^{{\bm k}_{\rm c}{\bm m}}_{n,i\alpha}~\delta{\dot m}_{\alpha}.
\end{align}
Here, $\delta \dot m_{\alpha}$ is the time derivative of $\delta m_\alpha$.
%The time derivative of the transverse components of the magnetization is explicitly written as $\delta \dot m_{\alpha}$, since the axis vector $\bm m_{0}$ is static.
% 
In the following, we omit the subscript ``c'' in ${\bm x}_{c}$ and ${\bm k}_{c}$ % ; denoting ${\bm x}_{c}$ as ${\bm x}$, and ${\bm k}_{c}$ as ${\bm k}$, for notational simplicity.
for the sake of brevity.

\subsection{Choice of distribution function}\label{subsec:notes_distribution}

The electric current is generally expressed as
\begin{align}\label{eq:current_general}
    {\bm j}(t) = -e \sum_{n}\int_{\bm k} f_{n}\, \dot{\bm x},
\end{align}
where $f_{n}$ is the electron distribution function for the $n$-th band and $\dot{\bm x}$ denotes the wave-packet velocity given in Eq.~(\ref{eq:wavepac_x_EOM}).
% , equation of motion for wave-packet center.
In general, the electron distribution depends on the trajectory of $\delta \bm m (t)$ and the nature of electron scattering.
% 
% 20260309修正
% \red{
% The latter arises may arise from microscopic processes such as impurity scattering and electron-phonon scattering, either elastic or inelastic, and is here represented phenomenologically by an effective relaxation time $\tau$.
% , for example, from disorder and phonons, and is often incorporated phenomenologically through a relaxation time $\tau$.
The latter is caused by impurity scattering as well as electron-phonon and electron-electron interactions, encompassing both elastic and inelastic scattering processes.
Such effects are well characterized by the relaxation time $\tau$, which measures how rapidly electrons lose the memory of their initial state.
% }
% 
% \blue{For the sake of simplicity}
Because we are interested in the intrinsic responses, 
we hereafter consider the case $\omega\tau\gg1$ ,i.e., the weak-scattering regime.
%** 
%** Special care is required with respect to the choice of the electron distribution function. 
% 
%** The appropriate choice should be determined by introducing the scattering processes characterized by the relaxation time $\tau$. 
% 
In the regime $\omega\tau \gg 1$ ($\tau \gg T=2\pi/\omega$), 
the driving period by magnetization precession is much shorter than the relaxation time. 
Therefore, the electron distribution does not change within one cycle~\cite{Harada2023}.
% In other words, we focus on a case in which 
% the intrinsic response dominates. 
% 

Assuming that the perturbation is switched on adiabatically, i.e. the Hamiltonian reduces to ${\hat{H}}^{(0)}$ in the initial state, 
the electron distribution function is 
taken as $f_{n}=f(\mathcal{E}^{(0)}_{n})$~\cite{Ishizuka2016,Ishizuka2017}, where $f$ is the Fermi-Dirac distribution function
\footnote{
In contrast, in the regime $\omega \tau \ll 1$, electrons undergo repeated scattering during one cycle,
therefore the extrinsic response dominates.
In this situation, the system does not retain memory of the past distribution, 
and therefore each moment can be regarded as a local equilibrium. 
Consequently, the appropriate distribution function is $f({\tilde \varepsilon}_{n})$, 
where ${\tilde \varepsilon}_{n}$ is the eigenvalues of the total Hamiltonian 
${\hat {H}}(t)={\hat {H}}^{(0)} + \delta m_\beta(t){\hat {H}}'_{\beta}$.
}. 
% 
% In contrast, in the opposite regime $\omega\tau \ll 1$, repeated scattering drives the system to local equilibrium at each moment, and the distribution must be evaluated from the instantaneous eigenvalues of ${\hat{\mathcal H}}(t)$. 
% (As our focus here is on intrinsic contributions, we restrict ourselves to the former case.)
% 
%
Here ${\mathcal E}^{(0)}_{n}$ is the energy of the $n$-th wave-packet in zeroth order of $\delta m$ and reduces to the unperturbed energy $\varepsilon^{(0)}_{n{\bm k}}(\bm m_0)$.
Therefore, in the situation described above, we should choose the electron distribution function as 
$f_{n}=f(\varepsilon^{(0)}_{n{\bm k}}(\bm m_0))$.

% \textcolor{red}{
% some explanations that ${\mathcal E}^{(0)}_{n}$ is reduced to bare band energy $\varepsilon_{n}$.
% }

Focusing on the intrinsic response up to the second order of the perturbation, 
% ${\mathcal O}(\delta m^2)$, 
%the expression for 
the current reads,
\begin{align}\label{eq:jPrecs_def}
     j_{i}(t ; {\bm m}_{0}) 
        &= -e \sum_{n} \int_{\bm k} f^{(0)}_{n}\,\dot{x}_{n,i} \notag \\
        &\approx -e \sum_{n} \int_{\bm k} f^{(0)}_{n} 
                 \left( 
                     \frac{1}{\hbar}\frac{\partial \mathcal{E}^{(2)}_{n}}{\partial k_{i}} 
                     + \sum_{\alpha}\Omega^{{\bm k}{\bm m}(1)}_{n,i\alpha}\,\delta \dot m_{\alpha}(t)
                 \right),
\end{align}
where $f^{(0)}_{n} = f(\varepsilon^{(0)}_{n{\bm k}}(\bm m_0))$.
Here, we explicitly denote the dependence of the precession axis ${\bm m}_{0}$ %to specify which axis 
as the SMF depends on it.
In the second line, we perform the expansion 
${\tilde{\mathcal E}}_n = \sum_{N} {\mathcal E}_n^{(N)}$ and 
${\tilde \Omega}^{{\bm k}{\bm m}}_{n,i\alpha} = \sum_{N} \Omega^{{\bm k}{\bm m}(N)}_{n,i\alpha}$,
where the terms with superscript $(N)$ are in %with respect to
the order of  $\delta m^N$.
% 
%The evaluation of 
Equation~(\ref{eq:jPrecs_def}) contains two contributions; 
the second order correction to the energy of the $n$-th wave-packet 
% $\mathcal{E}^{(2)}_{n}$ 
and the first order correction to the Berry curvature, 
% $\Omega^{{\bm k}{\bm m}(1)}_{n,i\alpha}$.
which will be evaluated in the following section.

\subsection{Ingredients of nonlinear spin-motive force}\label{subsec:current_expression}

%We first examine 
The second order correction to ${\tilde {\mathcal E}}_{n}$ in Eq.~(\ref{eq:jPrecs_def}), $\mathcal{E}^{(2)}_{n}$, is obtained by substituting Eq.~(\ref{eq:WavePac_nomalized}) into Eq.~(\ref{eq:WavePacEn_def}) 
(see Appendix.\ref{sec:Appendix.C}). 
%summing up with respect to the second order of perturbation~(see Appendix.\ref{sec:Appendix.C} for detailed derivation), 
The result reads,
\footnotesize
\begin{align}\label{eq:WavePac_En(2)}
     {\mathcal E}^{(2)}_{n} &=  J_{\rm ex}^{2}\sum_{\alpha,\beta}{\rm Re} \left[
                                               \sum_{l(\ne n)}  
                                               \frac{ \varepsilon^{(0)}_{nl,{\bm k}}({\bm m}_{0}) \sigma^{nl}_{\alpha} \sigma^{ln}_{\beta}}
                                                    {(\varepsilon^{(0)}_{nl,{\bm k}}({\bm m}_{0}))^2 - \hbar^{2}\omega^{2}}
                                         \right] \delta m_{\alpha}(t)\delta m_{\beta}(t) \notag \\
                            & 
                               +\hbar J_{\rm ex}^{2}\sum_{\alpha,\beta}{\rm Im} \left[
                                               \sum_{l(\ne n)} 
                                               \frac{ \sigma^{nl}_{\alpha} \sigma^{ln}_{\beta} }
                                                    { (\varepsilon^{(0)}_{nl,{\bm k}}({\bm m}_{0}))^2 - \hbar^{2}\omega^{2}}
                                         \right] \delta m_{\alpha}(t)\delta {\dot m}_{\beta}(t).
\end{align}
\normalsize

On the other hand, the first order correction to the Berry curvature reads
\begin{align}\label{eq:BC_1stPerturb}
     \Omega^{{\bm k}{\bm m}(1)}_{n, i\alpha} 
       &= \partial_{k_{i}} A^{{\bm m}(1)}_{n, \alpha} - \partial_{m_{\alpha}} A^{{\bm k}(1)}_{n, i}, 
\end{align}
where 
\begin{align}\label{eq:A(1)xi_1st_term}
      A^{{\bm k}(1)}_{n, i} 
      &= -i\bra{ u^{(1)}_{n{\bm k}} ({\bm m})} \partial_{k_{i}} \ket{u^{(0)}_{n{\bm k}} ({\bm m}_{0}) } \notag \\
      &\ \ \ \ 
         -i\bra{ u^{(0)}_{n{\bm k}} ({\bm m}_{0})} \partial_{k_{i}} \ket{ u^{(1)}_{n{\bm k}} ({\bm m})}
      \notag \\
      &= 2{\rm Re} \left[ -i\langle u^{(1)}_{n{\bm k}} ({\bm m}) | \partial_{k_{i}} |u^{(0)}_{n{\bm k}} ({\bm m}_{0}) \rangle  \right], 
\end{align} 
is the first order correction to the $\bm k$-space Berry connection for the $n$-th band state,
and $A^{{\bm m}(1)}_{n, \alpha}$ is defined in the same way
\footnote{In our calculations, the first-order correction to the Berry connection is evaluated by using the orthogonality relation between eigenstates 
$\braket{ u^{(1)}_{n\bm k}(\bm m) | u^{(0)}_{n\bm k} (\bm m_0)} = 0$
, which corresponds to fixing the gauge such that the Berry connection in the $\bm m$-space is zero.}, 
\begin{align}
    A^{{\bm m}(1)}_{n, \alpha} 
      &= -i\bra{ u^{(1)}_{n{\bm k}} ({\bm m})} \partial_{m_{\alpha}} \ket{u^{(0)}_{n{\bm k}} ({\bm m}_{0}) } \notag \\
      &\ \ \ \ 
         -i\bra{ u^{(0)}_{n{\bm k}} ({\bm m}_{0})} \partial_{m_{\alpha}} \ket{ u^{(1)}_{n{\bm k}} ({\bm m})}
      \notag \\
      &= 2{\rm Re} \left[ -i\langle u^{(1)}_{n{\bm k}} ({\bm m}) | \partial_{m_{\alpha}} |u^{(0)}_{n{\bm k}} ({\bm m}_{0}) \rangle  \right].
\end{align}
% 
% 
% \footnote{
% We here assume that the unperturbed Bloch state and the first-order perturbed Bloch state are orthogonal each other; $\langle u^{(1)}_n | u^{(0)}_n \rangle =0$, to derive the expression of Eq.(\ref{eq:A(1)xi_1st_term}).\\
% % 
% % これは
% % 
% % Since $\ket{u^{(1)}_n}$ can generally be expressed as $\ket{u^{(1)}_n} = \delta m_{\alpha}\partial_{\alpha}\ket{u^{(0)}_n} $ in perturbation theory,
% % (one can find the expression of $\partial_{\alpha}\ket{u^{(0)}_n}$ is exactly same as the Eq. (\ref{eq:Bloch_pertub}) with using, for instance, the Sternheimer equation.)
% % 
% This assumption corresponds to adopting the condition that ${\bm A}^{{\bm m}(0)}_n$ can be gauged zero in the $\bm m$-space globally.
% % 
% % 本研究では，磁化のダイナミクスの間にバンド交差が起こらないような状況に注目していること，また$\delta \bm m = 0$近傍の摂動理論を扱っているしている．これらのことから，ベリー接続${\bm A}^{{\bm m}(0)}_n$をゼロにゲージ固定するという仮定が，本研究の範囲内では，正当化できる．
% % 
% % The perturbative treatment in our study is carried out in the vicinity of $\delta{\bm m}=0$. 
% % Therefore, in summary, our calculation in this paper is conducted under the gauge such that ${\bm A}^{{\bm m}(0)}_n = 0$ in the region near the origin of the $\bm m$-space.
% }.
% 
Using Eq.~(\ref{eq:Bloch_pertub}), 
we find that the following relations hold for the first-order correction to $\bm k$-space and the $\bm m$-space Berry connection~(see also Appendix.\ref{sec:Appendix.A(1)derive} for the derivation): 
\begin{align}
     \label{eq:Am_(1)}
     A^{{\bm m}(1)}_{n, \alpha} &= - \sum_{\beta} \Bigl( {\overline \Omega}^{{\bm m}{\bm m}}_{n, \alpha \beta}\delta m_{ \beta}(t) 
                                -2 \hbar {\overline G}^{{\bm m}{\bm m}}_{n, \alpha \beta} \delta {\dot m}_{\beta}(t)\Bigr), \\
     \label{eq:Ak_(1)}
     A^{{\bm k}(1)}_{n, i}      &= - \sum_{\beta} \Bigl( {\overline \Omega}^{{\bm k}{\bm m}}_{n, i \beta}  \delta m_{\beta}(t)
                                -2 \hbar {\overline G}^{{\bm k}{\bm m}}_{n, i \beta}  \delta {\dot m}_{\beta}(t) \Bigr).
\end{align}
Here, we introduce the ``normalized'' Berry curvature ${\overline \Omega}^{{\bm \xi} {\bm \xi}'}_{n, \mu \nu}$ and Berry connection polarizability ${\overline G}^{{\bm \xi} {\bm \xi}'}_{n, \mu \nu}$ for ${\bm \xi} = {\bm k}, {\bm m}$. 
The concrete expressions for these quantities are given in Appendix~\ref{sec:Appendix.DefOfQGT}.
Substituting Eqs.~(\ref{eq:Am_(1)}) and (\ref{eq:Ak_(1)}) into Eq.~(\ref{eq:BC_1stPerturb}), 
we find the following expression for the first-order correction to the Berry curvature,
\begin{align}\label{eq:MBC_1stcorr}
     \Omega^{{\bm k}{\bm m}(1)}_{n,i\alpha} &= -\sum_{\beta}\left(
                                                      \partial_{k_{i}}{\overline \Omega}^{{\bm m}{\bm m}}_{n,\alpha \beta} 
                                                     -\partial_{m_{\alpha}}{\overline \Omega}^{{\bm k}{\bm m}}_{n,i\beta}
                                              \right)\delta m_{\beta}(t) \notag \\
                                          &
                                             -2\hbar\sum_{\beta}\left(
                                                      \partial_{k_{i}}     {\overline G}^{{\bm m}{\bm m}}_{n,\alpha \beta}  
                                                     -\partial_{m_{\alpha}}{\overline G}^{{\bm k}{\bm m}}_{n,  i    \beta}
                                               \right)\delta {\dot m}_{\beta}(t). 
%                                                     & \ \ \ \ \ 
%                                                        +  {\tilde \Omega}^{{\bm k}{\bm m}, n}_{i \beta} \partial_{m_{\alpha}}\delta m_{\beta}(t)
%                                                        +2 {\tilde {\mathcal G}}^{{\bm k}{\bm m}, n}_{i \beta} \partial_{m_{\alpha}}\delta {\dot m}_{\beta}(t)
\end{align}
%
% The detailed derivation of the perturbation correction for the Berry connection and curvature 
% are provided in Appendix.\ref{sec:Appendix.E}.
% 
% 
% Equations~(\ref{eq:WavePac_En(2)}) and (\ref{eq:MBC_1stcorr}), combining with Eq.~(\ref{eq:jPrecs_def}), 
% can be used to calculate the nonlinear spin motive force, as will be discussed in the following section. 

\subsection{ Expressions of nonlinear spin motive force }

% 式(16)および式(21)を非線形電流の表式(15)へ代入することで得られる電流は，３つのグループへ分類することができ，これらの表式が本論文のkey resultsである．
By substituting Eqs.~(\ref{eq:WavePac_En(2)}) and (\ref{eq:MBC_1stcorr}) 
into Eq.~(\ref{eq:jPrecs_def}), we find that the nonlinear spin-motive force has two contributions: 
direct current (DC) and second harmonic generation (SHG) components, 
\begin{align}\label{eq:jPrecs_catego}
     j_{i} (t;{\bm m}_{0}) = j_i(0;\bm m_0) + 2{\rm Re}[j_i(2\omega;\bm m_0) e^{-2i\omega t}].
     % j_{i} (t;{\bm m}_{0}) &= j^{\rm BC}_{i} (t;{\bm m}_{0}) + j^{\rm BCP}_{i} (t;{\bm m}_{0}) + j^{\rm surf}_{i} (t;{\bm m}_{0}).  \\
     %j_{i} (t;{\bm m}_{0}) &= j^{\rm DC}_{i} ({\bm m}_{0}) +  j^{\rm SHG}_{i} (t;{\bm m}_{0}).
\end{align}
%
% where in addition to the explicit time dependence, we also include $\bm m_0$, which specifies the axis around which the magnetization precesses, as an argument of the current.
% 
Here,
\begin{align}\label{eq:j_Fourier}
  j_i(\Omega;\bm m_0) &= \frac{1}{T}\int_0^{T} dt~e^{-i\Omega t} j_{i} (t;{\bm m}_{0}),
\end{align}
is the Fourier series of the nonlinear spin-motive force along $i$th direction at time $t$, $j_i(t;{\bm m}_{0})$.

\subsubsection{DC term}\label{subsubsec:DCterm}

% Instead of the time derivative of the magnetization $\delta {\dot {\bm m}}(t)$, 
% it would be better to express the results using the $\omega$-argument explicitly.
% In the following, we consider, for example, magnetization precession around the $x$-axis expressed as Eq.(\ref{eq:mag_vec}), 
% and explain the explicit forms of Eqs.~(\ref{eq:j(I)_rc}) to (\ref{eq:j(II)_SHG}) for the axis and their physical interpretations. %for each order of the current.

% \blue{
% Here, each term scales as a power of precession frequency as 
% $j^{\rm (I)}_{i} (t;{\bm m}_{0}) \propto {\mathcal O}(\omega)$, 
% $j^{\rm (II)}_{i} (t;{\bm m}_{0}) \propto {\mathcal O}(\omega^2)$, and 
% $j^{\rm (III)}_{i} (t;{\bm m}_{0}) \propto {\mathcal O}(\omega^0)$ 
% in the ``low'' frequency regime compared to a specific energy scale in the system, such as the width of the band gap.
% }

% \subsubsection{$\omega$-linear term}\label{subsubsec:omega_linear}

Firstly, let us focus on the DC term 
which corresponds to the $\omega=0$ component in Eq(\ref{eq:j_Fourier}).
% This term can be written as following form by using the Fourier component of derivation vector $\delta \bm m(\Omega)$ with respect to frequency,
%
Phenomenologically, the DC component reads
\begin{align}\label{eq:jDC_general}
     j_i(0;\bm m_0)   &= \sum_{\alpha, \beta}{\rm Re}~
                              % \Bigl[
                                   \sigma_{i;\alpha\beta}(0;\omega,-\omega)
                                   \delta m_{\alpha}(\omega)\delta m_{\beta}(-\omega),
                              % \Bigr].
                          % &= {\rm Re}
                          %     \Bigl[
                          %          \left( \sigma^{\rm BC}_{i;\alpha\beta}(0;\omega,-\omega)+\sigma^{\rm QM}_{i;\alpha\beta}(0;\omega,-\omega) \right)
                          %          \delta m_{\alpha}(\omega)\delta m_{\beta}(-\omega)
                          %                 \Bigr].
\end{align}
where $\sigma_{i;\alpha\beta}(0;\omega,-\omega)$ is the nonlinear conductivity.
Here, the deviation vector $\delta \bm m(\omega)$ is considered an external field.
The conductivity tensor consists of three contributions, 
$\sigma_{i;\alpha\beta}(0;\omega,-\omega) = \sigma^{\rm BC}_{i;\alpha\beta}(0;\omega,-\omega) + \sigma^{\rm BCP}_{i;\alpha\beta}(0;\omega,-\omega)$$ + \sigma^{\rm surf}_{i;\alpha\beta}(0;\omega,-\omega)$.
%, 
%and therefore the DC also has three origins; 
%$j_i(0;\bm m_0)  = j^{\rm BC}_i(0;\bm m_0) + j^{\rm BCP}_i(0;\bm m_0) +j^{\rm surf}_i(0;\bm m_0)$}.
Here, the first term is the Berry curvature-contribution, 
\begin{align}
    \label{eq:jDC_BC}
   &\sigma^{\rm BC}_{i;\alpha\beta}(0;\omega,-\omega) = 
   2ie\omega \sum_{n} \int_{\bm k} f^{(0)}_{n} 
            \Bigl(  
                  \partial_{k_{i}}     {\overline \Omega}^{{\bm m}{\bm m}}_{n, \alpha \beta} 
                - \partial_{m_{\alpha}}{\overline \Omega}^{{\bm k}{\bm m}}_{n,    i   \beta}
            \Bigr), 
\end{align}
and the second term is related to the Berry connection polarizability,
\begin{align}\label{eq:jDC_BCP}
   &\sigma^{\rm BCP}_{i;\alpha\beta}(0;\omega,-\omega) \notag \\
   &\ \ \ \ = 
   4ie\hbar\omega^2 \sum_{\gamma } \sum_{n} \int_{\bm k} f^{(0)}_{n} 
                                            \left(
                                            \partial_{k_{i}}     {\overline G}^{{\bm m}{\bm m}}_{n, \alpha \gamma}  
                                           -\partial_{m_{\alpha}}{\overline G}^{{\bm k}{\bm m}}_{n,   i    \gamma}
                                            \right) \epsilon_{\gamma\beta },
\end{align}
with $\epsilon_{\gamma \beta}$ being the Levi-Civita epsilon tensor. 
%\red{The interpretations of the above two terms will be discussed in the following section.}
The third term is a Fermi surface contribution, which reads, %takes the following forms
% includes the differential of the Fermi-Dirac distribution function with respect to the energy, and hence termed surface term. 
\begin{align}
  &\sigma^{\rm surf}_{i;\alpha\beta}(0;\omega,-\omega)\notag \\
  &=
  2e J_{\rm ex}^{2} 
                           \sum_{\substack{n,l\\(n \ne l)}}\int_{\bm k} f^{(0)'}_{n} 
       {\rm Re} 
         \left[
               \frac{ \varepsilon^{(0)}_{nl,{\bm k}}({\bm m}_0)~
                      v^{nn}_{i}\sigma^{nl}_{\alpha} \sigma^{ln}_{\beta} }
                    { (\varepsilon^{(0)}_{nl,{\bm k}}({\bm m}_0))^2 - \hbar^2 \omega^{2}}
         \right]
\end{align}
where $f^{(0)'}_{n}$ is the energy derivative of the Fermi distribution function.

In an insulator, $\sigma^{\rm surf}_{i;\alpha\beta}(0;\omega,-\omega)$ does not contribute to the nonlinear spin-motive force.
On the other hand, when the Fermi level crosses a band, this term seemingly gives a nonzero contribution. In the present study, we consider the limit in which electronic relaxation is much slower than the time scale of the magnetization dynamics. In realistic situations, however, a finite relaxation time is present, and this term is expected to exhibit a transient-current-like behavior, decaying within the scale of the relaxation time.

% Firstly, let us focus on $j^{\rm BC}_{i} (t;{\bm m}_0)$.
% This term originates from the second term of the wave packet energy correction in Eq.(\ref{eq:WavePac_En(2)}) together with the first term of the Berry curvature correction in Eq.(\ref{eq:BC_1stPerturb}), 
% and it scales with $\delta {\dot m}_{\alpha} \delta m_{\beta}$.
% Combining with Eq.(\ref{eq:mag_vec}), 
% the involvement of a single time derivative of $\delta {\bm m}$ implies that $j^{\rm BC}_{i} (t)$ scales linearly with $\omega$, 
% $j^{\rm BC}_{i} (t;{\bm m}_{0}) \propto {\mathcal O}(\omega)$ 
% (The precise $\omega$-dependence, including the $\omega$-argument of the normalized Berry curvature, is derived in Appendix. \ref{sec:Appendix.G}).

\subsubsection{SHG term}
Next, we turn to the SHG term
which is an AC response with frequency $2\omega$ in Eq(\ref{eq:j_Fourier}).
This term can be written as follows, 
\begin{align}\label{eq:jSHG_general}
      j_i(2\omega;\bm m_0) &= \sum_{\alpha, \beta}%{\rm Re}~
                                  \sigma_{i;\alpha\beta}(2\omega;\omega,\omega)~
                                  \delta m_{\alpha}(\omega)\delta m_{\beta}(\omega).
                                  % e^{-i2\omega t}
                                        % &= {\rm Re}
                                        %   \Bigl[
                                        %      \left( \sigma^{\rm BC}_{i;\alpha\beta}(2\omega;\omega,\omega)+\sigma^{\rm QM}_{i;\alpha\beta}(2\omega;\omega,\omega) \right)
                                        %       \delta m_{\alpha}(\omega)\delta m_{\beta}(\omega)
                                        %       e^{-i2\omega t}
                                        %   \Bigr].
\end{align}
Similar to the DC term, the conductivity tensor consists of three contributions, 
$\sigma_{i;\alpha\beta}(2\omega;\omega,\omega) = \sigma^{\rm BC}_{i;\alpha\beta}(2\omega;\omega,\omega) + \sigma^{\rm BCP}_{i;\alpha\beta}(2\omega;\omega,\omega)$$ + \sigma^{\rm surf}_{i;\alpha\beta}(2\omega;\omega,\omega)$.
%, so that 
%$j_i(2\omega;\bm m_0)  = j^{\rm BC}_i(2\omega;\bm m_0) + j^{\rm BCP}_i(2\omega;\bm m_0) +j^{\rm surf}_i(2\omega;\bm m_0)$
%}.
%Their explicit expressions are as follows, 
Here,
\begin{align}
    \label{eq:jSHG_BC}
   &\sigma^{\rm BC}_{i;\alpha\beta}(2\omega;\omega,\omega) = 
   -ie\omega \sum_{n} \int_{\bm k} f^{(0)}_{n} 
             \partial_{m_{\alpha}}{\overline \Omega}^{{\bm k}{\bm m}}_{n,    i   \beta}, \\
    \label{eq:jSHG_BCP}
   &\sigma^{\rm BCP}_{i;\alpha\beta}(2\omega;\omega,\omega) \notag \\
   &=  
   -2ie\hbar\omega^2 \sum_{\gamma } \sum_{n} \int_{\bm k} f^{(0)}_{n} 
                \left(
                      \partial_{k_{i}}     {\overline G}^{{\bm m}{\bm m}}_{n, \alpha \gamma}  
                     -\partial_{m_{\alpha}}{\overline G}^{{\bm k}{\bm m}}_{n,   i    \gamma}
                \right) \epsilon_{\gamma\beta },
\end{align}
\begin{align}
  &\sigma^{\rm surf}_{i;\alpha\beta}(2\omega;\omega,\omega)\notag \\
  &=
  e J_{\rm ex}^{2} 
                           \sum_{\substack{n,l\\(n \ne l)}}\int_{\bm k} f^{(0)'}_{n} 
       {\rm Re} 
         \left[
               \frac{ \varepsilon^{(0)}_{nl,{\bm k}}({\bm m}_0)~
                      v^{nn}_{i}\sigma^{nl}_{\alpha} \sigma^{ln}_{\beta} }
                    { (\varepsilon^{(0)}_{nl,{\bm k}}({\bm m}_0))^2 - \hbar^2 \omega^{2}}
         \right]
\end{align}
are the Berry curvature, Berry connection polarizability, and Fermi surface contributions, respectively.

% 
% DCの方は，フーリエ成分を考えるときに\delta m_\alpha(\omega)\delta m_\beta(-\omega)+\delta m_\alpha(-\omega)\delta m_\beta(\omega)がz+z*みたくなるから「2」Reが出る．だけどSHGの方はそういうのは無いので2倍は出てこない．
% 

\subsubsection{Interpretations of results}\label{subsubsec:physical_interp}

In previous studies, currents arising from the Berry curvature have been investigated primarily in the linear regime with respect to $\delta {\bm m}(t)$ in the adiabatic regime \cite{Freimuth2015, JTang2024, Manchon2024}.
It arises from using the unperturbed  Berry curvature ${\Omega}^{\bm k\bm m (0)}_{n}$ as $\tilde{\Omega}^{\bm k\bm m}_{n}$ in Eq. (\ref{eq:wavepac_x_EOM}).
% }
% (just substituting the unperturbed Berry curvature as a wave-packet velocity in Eq.(\ref{eq:current_general})).
It is nothing but the displacement current for $\delta\bm m(t)$ which is periodic in time.
% That current, known as \green{the} Thouless pump, is finite even in the insulating energy region if the system has a nontrivial topology \cite{KS_V, Freimuth2015, JTang2024}.
% 
In our formalism, the contribution from $\delta\bm m(t)$ in the perturbation theory tunes the Bloch wavefunction, 
and hence the expression of the Berry curvature. % in the Thouless pump-formula.
This leads to Eqs. (\ref{eq:jDC_BC}), (\ref{eq:jSHG_BC}) and can be interpreted as the extension of the formula in the adiabatic regime.
The direct current induced by the Berry curvature contribution, $\sigma^{\mathrm{BC}}(0;\omega,-\omega)$, is proportional to $m_\alpha \dot{m}_\beta - m_\beta \dot{m}_\alpha$, as $m_\alpha \dot{m}_\beta + m_\beta \dot{m}_\alpha$ term vanishes after taking the time integral.
As a consequence, the induced current is proportional to the area swept by $\bm m(t)$.
To see this, we note that the time dependence of this contribution arises from terms proportional to the product $m_\alpha \dot{m}_\beta$, whose antisymmetric component, $\bm{m} \times \dot{\bm{m}}$, corresponds to $\sigma^{\mathrm{BC}}(0;\omega,-\omega)$.
In a precessional motion, the magnetization traces a closed trajectory in $\bm{m}$-space, and $\bm{m} \times \dot{\bm{m}}$ corresponds to the area enclosed by this curve.
Thus, when the magnetization dynamics forms such a closed loop, the $\sigma^{\mathrm{BC}}(0;\omega,-\omega)$ contribution is proportional to the area.

By contrast, the symmetric component, $m_\alpha \dot{m}_\beta + m_\beta \dot{m}_\alpha$, %one can readily verify that substituting the explicit time dependence of the magnetization components 
yields terms proportional to $\cos 2\omega t$ and $\sin 2\omega t$, i.e., it contributes to the SHG through $\sigma^{\mathrm{BC}}(2\omega;\omega,\omega)$ term. 
Mathematically, this type of product can be written as a total time derivative, $\frac{d}{dt}(m_\alpha m_\beta)$. 
Upon integrating the current over time, one obtains a contribution of the form $m_\alpha(t_{\mathrm{end}}) m_\beta(t_{\mathrm{end}}) - m_\alpha(t_{\mathrm{initial}}) m_\beta(t_{\mathrm{initial}})$, where $t_{\mathrm{initial/end}}$ is beginning or end of the time integral. 
This term vanishes when the magnetization trajectory is closed as discussed above.
On the other hand, it yields a finite value for an open trajectory, corresponding to the oscillatory components at frequency $2\omega$.

In addition to the Berry curvature contribution, 
we find contributions arising from a non-adiabatic effect  (Eqs. (\ref{eq:jDC_BCP}) and (\ref{eq:jSHG_BCP})). %in nonadiabatic regime, which does not appear in previous studies.
These contributions are related to the normalized Berry connection polarizability. %, represent the nonadiabatic correction for the adiabatic current.
The dynamical correction proportional to $\delta \dot{\bm m}(t)$ in perturbation theory is responsible for the virtual inter-band mixing for the Bloch wavefunction, which is presented in Eqs. (\ref{eq:WavePac_En(2)}) and (\ref{eq:MBC_1stcorr}).
This leads to the correction beyond the adiabatic regime $\sim {\mathcal O}(\omega)$, which can be confirmed by the factor $\sim {\mathcal O}(\omega^2)$ in Eqs. (\ref{eq:jDC_BCP}) and (\ref{eq:jSHG_BCP}).
The above discussion is analogous to the one for the intrinsic nonlinear Hall effect \cite{YGao2014, Sodemann2015}.
In that situation, the electric field, the temporal dynamics of the vector potential, mixes the Bloch wavefunctions of different bands, which leads to the appearance of the the Berry connection polarizability in the $\bm k$-space.
In our situation, the Berry connection polarizability in the $(\bm k, \bm m)$-mixed space arises from the temporal dynamics of the magnetization.

% \red{The physical interpretation of BCP-term is ...
% the correction such that the electric current (polarization ?) can be generated even if the contour in the periodic parameter space does not close. 
% The analogy with the shift current ; nonlinear current response for the time-varying electric field ?
% }

The current arising from the Berry connection polarizability appears through the product $\dot{m}_\alpha \dot{m}_\beta$. 
For a precessing magnetization, this product explicitly takes the form $(1 + \cos 2\omega t)$ for $\alpha = \beta$, or $\sin 2\omega t$ for $\alpha \neq \beta$. 
The constant factor 1 in the former case, which corresponds to the time-averaged magnitude of the magnetization dynamics, gives rise to $\sigma^{\mathrm{BCP}}(0;\omega,-\omega)$, while the time-dependent deviation from it represents $\sigma^{\mathrm{BCP}}(2\omega;\omega,\omega)$. 
The time dependence of $\sigma^{\mathrm{BCP}}$ is governed not so much by a geometric interpretation in terms of whether the magnetization traces a closed loop, but rather by the temporal dynamics of the magnetization itself.

\section{Luttinger model}\label{sec:modelcalc}
%\section{Model calculation}\label{sec:modelcalc}

\subsection{Hamiltonian}
%\subsection{Effective model}

%===== FIG =====%
\begin{figure}[t]
   \includegraphics[width=1.0\hsize]{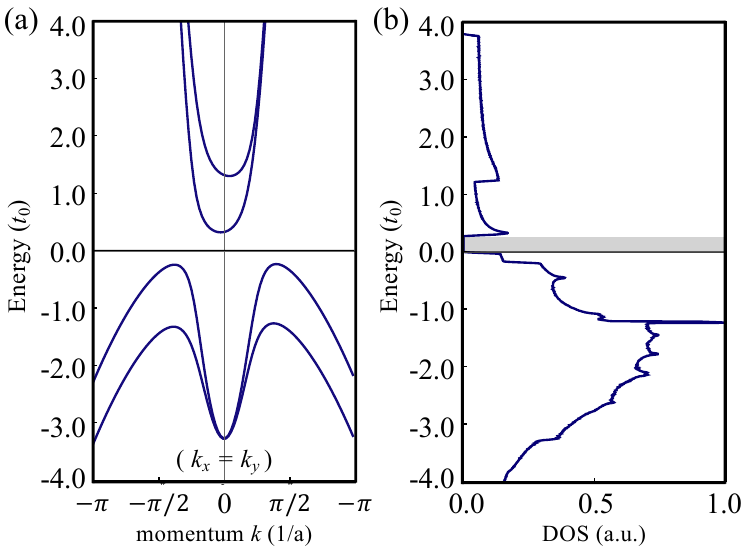} %Fig1_forArticle_260310.pdf
\caption{
(a) The band structure of $\hat {\cal H}_{0,\bm k}$ on the $k_{x} = k_{y}$ line.
(b) Density of states. The gray shade represents energy region where band is gapped.
The model parameters are set as 
$\gamma_1 = 3.0t_{0}a^2$, $\gamma_2 = 2.0t_{0}a^2$, $\gamma_3 = 1.5t_{0}a^2$, $\Delta = 2.0t_{0}$, 
$E_{\rm R} = 0.08t_{0}a$, $E_{\rm D} = 0.3t_{0}a$, and $J_{\rm c} = 0.5t_{0}$, 
where $t_{0}$ is the energy unit and $a$ is the lattice constant.
The direction of the equilibrium magnetization is set along the $x$-direction ; $\bm m_0 = \bm e_x$.
} 
\label{fig:band_DOS}
\end{figure}
%===============%
Now, we demonstrate the implementation of our theory in concrete materials.
As an example for demonstrating the quantum geometric effect, we consider magnetic precession in a two-dimensional magnetic semiconductor. 
The static part of the Hamiltonian reads
\begin{align}                                                      
     \label{eq:Ham_tot}
    &{\hat{\mathcal H}}^{(0)}_{{\bm k}} = {\hat{\mathcal H}}_{\rm L} + {\hat{\mathcal H}}_{\rm SOC} + {\hat{\mathcal H}}_{\rm exc}, \\
     \label{eq:Ham_L}
     &{\hat{\mathcal H}}_{\rm L}  = \frac{\gamma_1}{2}{\bm k}^{2} 
                                  + \gamma_{2}\left( \frac{5}{4}{\bm k}^{2} 
                                                   + \sum_{a=x,y}k_{a}^{2} {\hat S}^{2}_{a} \right) \notag \\
                                 &\ \ \ \ \ \ \ \ \ \ \ \ \ \ \ 
                                  - \gamma_{3} k_{x}k_{y}\{ {\hat S}_{x}, {\hat S}_{y} \}
                                  - \Delta \left( {\hat S}^{2}_{z} - \frac{5}{4} \right), \\
    \label{eq:Ham_SOC}
    &{\hat{\mathcal H}}_{\rm SOC}  =  E_{\rm R} \left( k_{x}{\hat S}_{y}-k_{y}{\hat S}_{x} \right)
                                    + E_{\rm D} \left( k_{x}{\hat S}_{x}-k_{y}{\hat S}_{y} \right), \\
    \label{eq:Ham_exc}
    &{\hat{\mathcal H}}_{\rm exc}  = - J_{\rm ex} {\bm m}_{0} \cdot {\hat {\bm S}},
\end{align}
where ${\bm k}=(k_{x}, k_{y})$, and ${\hat {\bm S}} = ({\hat S}_{x}, {\hat S}_{y}, {\hat S}_{z})$ is the spin-$3/2$ operator. %\cite{SMurakami2004}.
% The explicit form of the matrices ${\hat {\bm S}}$ is given in Appendix~\ref{sec:Appendix.H}.
% 
The first term in Eq.~(\ref{eq:Ham_tot}) is a variant of the Luttinger Hamiltonian \cite{Luttinger1956} 
where $\gamma_{1,2,3}$ are the Luttinger parameters %characterize the overall shape of the band structure whereas 
and $\Delta$ is the out of plane anisotropy.
The first and second terms in Eq.~(\ref{eq:Ham_SOC}) are respectively the Rashba- and Dresselhaus-like spin-orbit couplings, which generally exist in a noncentrosymmetric system. 
%comes from the confinement of our system into a two-dimensional shape, such as heterostructure.
% 
Finally, the third term, Eq.~(\ref{eq:Ham_exc}), is the exchange coupling between the electrons and the local moment. 
% \green{the direction of static magnetic field}. %aligned with the axis of the precession. 
%\green{The dispersion and the density of states (DOS) of ${\hat{\mathcal H}}_{0,{\bm k}}$ are shown in Fig.~\ref{fig:band_DOS}.}
Throughout this section, we set %In our analysis, 
% we set the lattice constant $a=1$, except for the concrete estimation of the electric current (see below). 
%each parameter is set as 
$\gamma_1 = 3.0t_{0}a^2$, $\gamma_2 = 2.0t_{0}a^2$, $\gamma_3 = 1.5t_{0}a^2$, $\Delta = 2.0t_{0}$, 
$E_{\rm R} = 0.08t_{0}a$, $E_{\rm D} = 0.3t_{0}a$, $J_{\rm ex} = 0.5t_{0}$, 
where $t_{0}$ is the energy unit and $a$ is the lattice constant; 
%In the following, 
we set %$t_{0} = 10~{\rm meV}$ and 
$a=1$, 
except for the order estimation of the electric current.  

Figure~\ref{fig:band_DOS} shows the dispersion along the $k_x = k_y$ line and the density of states (DOS) of ${\hat{\mathcal H}}_{0,{\bm k}}$ where the axis vector $\bm m_0$ is aligned with the $x$-direction.
%and density of states~(DOS).
% 
The electronic band consists of two valence and two conduction bands; it is an indirect semiconductor with the indirect gap of $\Delta\simeq 0.3t_0$ at around $E=1.2t_0$.
The gap is shown as a gray shade in Fig.~\ref{fig:band_DOS}(b).
% 
%We note that the minimum gap is not placed on the high symmetry line such as the $k_x = k_y$ line. 
%One can easily find that this system is a ferromagnetic semiconductor with gap $\approx 0.3 t_0$ 
%which is depicted as a gray shade region in Fig.\ref{fig:band_DOS}~(b).

For the effective model described above, 
we consider the situation where magnetization precesses around the $x$-axis and 
the nonlinear SMF generated along the $x$-axis.
Hence, we now evaluate Eqs.(\ref{eq:jDC_BC}), (\ref{eq:jDC_BCP}), (\ref{eq:jSHG_BC}), and (\ref{eq:jSHG_BCP}) for $i=x$ numerically.

%===== FIG =====%
\begin{figure*}[t]
   \includegraphics[width=1.0\hsize]{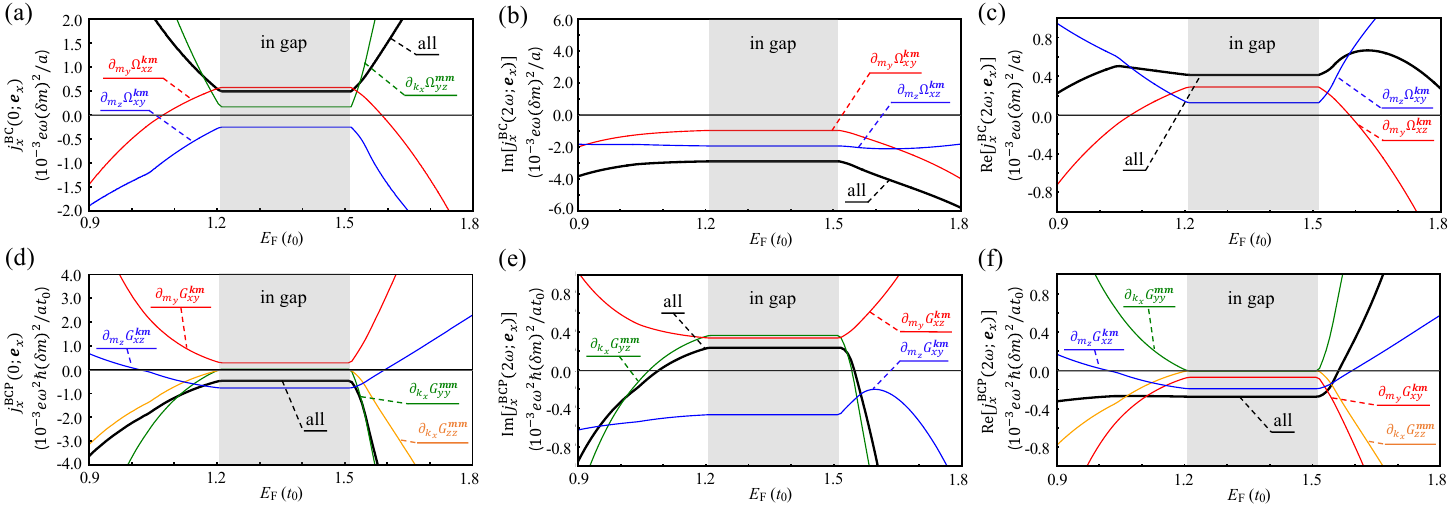} %Fig2_forArticle_revised_260120.pdf
\caption{
Calculation results of the electric current along the $x$-axis under the magnetization precession around the axis $\bm m_0 = \bm e_x$.
Upper (lower) panels show the response coefficients arising from the Berry curvature (Berry connection polarizability). 
Panels (a,d) show the direct-current component, while (b,e) and (c,f) show the $\sin 2\omega t$ and $\cos 2\omega t$ components of second-harmonic generation, respectively.
The gray shade represents the energy region where the band is gapped.
The colored lines in each panel represent the contributions of the geometric quantities that arise when one expands the summation for the magnetization direction, $\alpha$ and $\beta$, 
in the expression of the electric current, Eqs. (\ref{eq:jDC_general}) and (\ref{eq:jSHG_general}), explicitly.
} 
\label{fig:jSHGsin_x;yz}
\end{figure*}
%===============%

\subsection{Results of numerical calculation}\label{seubsec:numerical}

% For the effective model of DMS described above, 
% we consider the situation where magnetization precesses around the $x$-axis and 
% the nonlinear SMF generated along the $x$-axis.
% % 
% We now evaluate Eqs.(\ref{eq:j(I)_rc_mx}) to (\ref{eq:j(II)_cos_mx}) for $i=x$ numerically. 
% %

Figures \ref{fig:jSHGsin_x;yz}(a), (b), and (c) show the results of the numerical calculation for the $\omega$-linear contributions to the dc current and SHG
(Eqs.(\ref{eq:jDC_BC}) and (\ref{eq:jSHG_BC})) as a function of the Fermi energy $E_F$.
The gray shade represents the region of $E_F$ in which the Fermi level is in the band gap. 
We plot the specific energy region around the gap for the sake of visibility.
In Fig. \ref{fig:jSHGsin_x;yz}(a), the SMF %exhibits a finite and constant value 
occurs even when the Fermi level is in the energy gap, i.e., without mobile carriers. 
The same behavior is also observed in Figs. \ref{fig:jSHGsin_x;yz}(b) and \ref{fig:jSHGsin_x;yz}(c).
Since this region corresponds to the insulating regime, there are no conduction electrons contributing to the current.
With taking into account the appearance of the plateau in insulating regime, 
Figs.~\ref{fig:jSHGsin_x;yz}(a)-\ref{fig:jSHGsin_x;yz}(c) indicates that the $\omega$-linear term behaves in a displacement current-like manner.
Figures \ref{fig:jSHGsin_x;yz}(d), (e), and (f) show the results of the numerical calculation of the $\omega$-quadratic SMF as a function of the Fermi energy 
(Eqs.(\ref{eq:jDC_BCP}) and (\ref{eq:jSHG_BCP})).
These numerical results also exhibit a plateau with a finite value inside the gap.
% \blue{These plateau indicate that $\omega$-quadratic current behaves in a displacement current ...}
% \red{However, it should be noted that its magnitude is smaller by about three orders of magnitude ($10^{-3}$) compared to that of the $\omega$-linear current} \blue{in our setting}. 
% 

In addition, we find that, at least for the present model, the formations of the current plateau inside the gap in Fig.\ref{fig:jSHGsin_x;yz} are predominantly governed by ${\overline \Omega}^{\bm k\bm m}_n$ or ${\overline G}^{{\bm k}{\bm m}}_n$, rather than by ${\overline \Omega}^{\bm m\bm m}_n$ or ${\overline G}^{{\bm m}{\bm m}}_n$. 
This suggests that the magnetic semiconductor described by Eq.(\ref{eq:Ham_tot}) hosts the nontrivial quantum geometry in the $(\bm k, \bm m)$-mixed space.
On the other hand, this model does not have a nontrivial distribution of the Berry curvature $\Omega^{\bm k\bm k}_{n}$ and, therefore, shows neither the anomalous Hall effect nor the finite Chern number.
In that sense, this model is a ``trivial'' model in the context of the conventional $\bm k$ space-quantum geometry.
Our calculations indicate that even magnetic materials which are previously regarded as quantum-geometrically trivial can exhibit nontrivial spin-charge conversions from the perspective of the $(\bm k, \bm m)$-mixed space-quantum geometry.

\begin{table}[htbp]
    \centering
% \begin{center}
\renewcommand{\arraystretch}{1.7}
\begin{tabular}{||l|c||l|c||} \hline
     &  $({\rm nA/cm})$ &  & $({\rm nA/cm})$\\ \hline
   (a) : $j^{\rm BC}_x(0;{\bm e}_x)$ & $3.62$ & (d) : $j^{\rm BCP}_x(0;{\bm e}_x)$ & $-1.01 $ \\ \hline
   (b) : ${\rm Im}[j^{\rm BC}_x(2\omega;{\bm e}_x)]$ & $-21.1 $ & (e) : ${\rm Im}[j^{\rm BCP}_x(2\omega;{\bm e}_x)]$ & $0.517 $ \\ \hline
   (c) : ${\rm Re}[j^{\rm BC}_x(2\omega;{\bm e}_x)]$ & $3.02 $ & (f) : ${\rm Re}[j^{\rm BCP}_x(2\omega;{\bm e}_x)]$ & $-0.601 $ \\ \hline
 \end{tabular}
 \renewcommand{\arraystretch}{1.0}
% \end{center}
\caption{
% \red{
Summary of the values for the electric currents converted into the unit of the current density $({\rm nA/cm})$.
In the table, (a)-(f) denote the electric currents inside the gap in the respective panels of Fig. \ref{fig:jSHGsin_x;yz}.
We assume that the frequency of the magnetization precession is $\omega = 4.56 \times 10^1$ GHz.
% }
}
\label{table:estimate}
\end{table}

At the end of this section, let us estimate the numerical values of the current in Fig.~\ref{fig:jSHGsin_x;yz}. %by converting the unit to that of the current density, A/cm.
To this end, we set the lattice constant to $a=10^{-7}$ cm, corresponding to the order of angstroms. 
The frequency is chosen as $\hbar\omega=0.03$ meV, which %is of the same order as microwave energy, and hence
corresponds to $\omega = 4.56 \times 10^1$ GHz.
The amplitude of the magnetization precession is set to $\delta m = 0.01$, to ensure the validity of the perturbative treatment. 
The unit of energy was taken as $t_0=10$ meV. 
With these parameters, the numerically obtained energy gap, indicated by the gray-shaded region, becomes approximately $3$ meV, two orders of magnitude larger than $\hbar\omega$.
Thus, justifying the adiabatic approximation.
The current values converted under these parameter settings are summarized in Table \ref{table:estimate}, 
which are in the order of $10^{-1} \sim 10^1$ nA/cm.
Therefore, the current induced by our mechanism can be detected on the usual current ammeter.
% 
% 

% Note that this value is comparable to ...
% already measured in  ferromagnetic systems. 
% It should be detectable in experiment, e.g., by AAA or BBB ...
% \blue{Experimentally, a realistic approach to detect these nA/cm-order current is contactless near-field magnetometry, such as $\rm \mu$SQUID or NV-center AC magnetometry, or  followed by Biot-Savart inversion.}
% Direct electrical detection is also possible, but it requires measures such as IQ heterodyne detection or bridge configurations to suppress systematic artifacts from spin rectification and thermoelectric effects.

% \textcolor{red}{($\cdots$ 
% the results for another model parameters~(zero Rashba, making Dresselhaus SOC stronger, varying Luttinger parameters...), 
% approaches or strategies to enhance the nonlinear current, etc)}

% \section{V. Discussion}\label{sec:discussion}
% \textcolor{red}{
% \begin{itemize}
%     \item frequency dependencies, which is leading term ?
%     \item top. SOT and ISOT : Onsager-like understanding, Intrinsic second-order SOT and WHAT, the point of view from whether the geometric quantities (mBC, QM, BCP) have {\it nontrivial} distribution
%     \item material consideration : MnBiTe, MWSM et al..., nontrivial distribution of mBC, QM, and BCP
% \end{itemize}
% }

\section{Discussion and Conclusion}\label{sec:discuss&concl}

% \red{Discussion for the approaches or strategies to enhance the nonlinear current and material consideration.}

% 簡単に論文のまとめ
In this work, we theoretically proposed the intrinsic nonlinear SMF which arises from the quantum geometry in the $({\bm k}, {\bm m})$-mixed space.
We formulated the electron dynamics in the ferromagnets by the semiclassical wave-packet formalism up to the second order of the amplitude of the time-varying magnetization.
As a result, we found that the nonlinear SMF has two contributions; the DC term and the SHG term. 
Both of them can be written in terms of the Berry curvature and the quantum metric in the $({\bm k}, {\bm m})$-mixed.
To illustrate the significance of the nonlinear SMF, 
we performed numerical calculations for a model of two-dimensional magnetic semiconductors. 
As a result, we found that, even in the insulating regime, the nonlinear SMF shows a finite value, and that value can be detected by the usual current ammeter.
% These findings suggest that 
% our mechanism serves as a new operating principle for an AC-to-DC converter, governed solely by the electron's geometric property in magnetic materials.

% 
% The current originating from the Berry curvature has the physical interpretation that it is the correction for the Thouless pump driven by $\delta \bm m (t)$, 
% while the one originating from the quantum metric has the physical interpretation \blue{that ...}
% Such a interpretation is confirmed by the model calculation for the two dimensional magnetic semiconductor where the induced current shows the finite plateau in insulating energy region.

% nonlinear SMFを大きくする指針はあるか
The nonlinear SMF we propose is expected to exist broadly in various materials, regardless of whether they are metallic, insulating, or semiconducting.
As strategies for observing or enhancing the nonlinear SMF, we first suggest exploring materials or junction systems in which the $({\bm k}, {\bm m})$-space quantum geometry is already known to exhibit a nontrivial structure, as inferred from studies on current-induced spin-orbit torques \cite{HKurebayashi2014, Freimuth2014, Hanke2017, CXiao2022_2ndSOT}. 
Second, one may focus on materials with topologically nontrivial band structures or with band degeneracy points. In this sense, magnetic Weyl semimetals provide a promising platform \cite{Meguro2025} and constitute the subject of our future studies. 
More recently, MnBi$_2$Te$_4$ has also been proposed as an intriguing candidate system \cite{JTang2024, JXQiu2025}.

% 外因性SMFの存在について言及
% In this study, we have focused on the intrinsic response, but extrinsic responses arising from the Fermi surface and the associated relaxation time must also be taken into account. 
% % 
% In principle, such extrinsic effects can be incorporated within the our theoretical framework by appropriately solving the electron distribution function in Eq. (\ref{eq:current_general}). 
% The resulting currents can, in principle, be distinguished by tuning the Fermi level into the gap via doping, or through their characteristic scaling with the relaxation time, similarly to the case of voltage-driven nonlinear current responses.

% ゲージ固定の話
% 本研究では，ベリー接続の一次補正を計算する際に，固有状態同士の内積がゼロであるという関係式を用いているが，これは実質的には磁化ベクトル空間（m空間）におけるベリー接続のゲージをゼロに固定している状況に対応している．このようなゲージ条件の下では，線形応答領域のThouless pumpは生じない．本研究では非線形応答を主題としており，Thouless pumpが生じない状況を対象とすることで，非線形応答が支配的に現れる場合を考えた．一方で，上述のMWSMのように，非自明なバンド構造をもつ物質では，一般にm空間ベリー接続のゲージを大域的にゼロに固定することはできないと考えられる．そのため，そのような場合には，上記の内積ゼロの関係式に依存せずにベリー接続の一次補正を評価する必要が生じる．このより一般的な取り扱いについては，すでに別途計算を進めており，現在[in preparation]としてまとめている．

% In our calculations, the first-order correction to the Berry connection is evaluated by using the orthogonality relation between eigenstates 
% $\braket{ u^{(1)}_{n\bm k}(\bm m) | u^{(0)}_{n\bm k} (\bm m_0)} = 0$
% , which corresponds to fixing the gauge such that the Berry connection in the $\bm m$-space vanishes. 
% 
In our calculations, we assume that we can fix the gauge of the Berry connection in the $\bm m$-space to zero globally.
Under this gauge condition, linear-response current, Thouless pump, does not arise.
Since the main focus of this study is on nonlinear response, we intentionally restrict ourselves to situations where the Thouless pump is absent, and hence the nonlinear contribution becomes dominant. 
On the other hand, for materials with nontrivial band structures where 
% such as the magnetic Weyl semimetals mentioned above, one generally cannot choose
a gauge cannot sets the $\bm m$-space Berry connection to zero globally, 
% 
% In such cases, 
it becomes necessary to evaluate the first-order correction to the Berry connection without relying on the above orthogonality condition. 
The results for such cases will be reported elsewhere \footnote{
T. Meguro, H. Ishizuka, and K. Nomura, in preparation.
}.

% 最後に磁化->電気的応答における量子幾何の重要性を強調
Finally, we emphasize the crucial role of quantum geometry in the electric response driven by magnetization dynamics.
The usefulness of the $({\bm k}, {\bm m})$-mixed space quantum geometry has been discussed primarily in the context of current-induced spin-orbit torques which is the charge-to-spin conversion
\cite{Freimuth2014,CXiao2021,CXiao2022_2ndSOT}. 
% 
% \red{
Although preceding studies have shown that the reciprocal process, SMF, is written by the Berry curvature in mixed space, it is limited to the linear response regime and hence AC.
It was also unknown how the quantum metric affects the conversion process from the magnetic dynamics to electric current.
Our study demonstrates that the SMF in the nonlinear regime can convert injected frequency into DC rectification and SHG, and their geometric origins are distinguished by their frequency scaling.
This suggests that 
our mechanism serves as a new operating principle for an AC-to-DC converter, governed solely by the electron's geometric property in magnetic materials.
% }
% 
Our finding not only forms the basis for exploring rich intrinsic nonlinear electric responses in dynamical magnetic systems, but also serves as a building block for the emerging field of nonlinear spintronics.

\section*{Acknowledgments}

The authors thank Naoto Nagaosa, Shuichi Murakami, and Takahiro Morimoto for fruitful discussions.
This work was supported by JST SPRING, Grant No. JPMJSP2136 (T.M.), 
JSPS KAKENHI, Grant No. JP25H00841 (H.I.), JP25H01250 (K.N.),
and JST PRESTO, Grant No. JPMJPR2452 (H.I.).

\bibliography{ref}

%%%%%%%%%%%%%%%%%%%%%%%%%%%%%%%%%%%%%%%%%%%%%%%%%%%%%%%%%%%%%%%%%%%%%%%%%%%%%%%%%%%%%%%%
%%%%%%%%%%%%%%%%%%%%%%%%%%%%%% Appendix zone %%%%%%%%%%%%%%%%%%%%%%%%%%%%%%%%%%%%%%%%%%%
%%%%%%%%%%%%%%%%%%%%%%%%%%%%%%%%%%%%%%%%%%%%%%%%%%%%%%%%%%%%%%%%%%%%%%%%%%%%%%%%%%%%%%%%

\section*{Appendix}\label{sec:Appendix}
\appendix
\section{Derivation of the correction to the wave function }\label{sec:Appendix.A}

Firstly, let us derive the differential equation for the coefficients $C_{l\bm k}(\bm x_c, \bm k_c, t)$.
Hereafter, we omit  $(\bm x_c, \bm k_c, t)$-dependence of the coefficients for notational simplicity.
The wave-packet follows the time-dependent Schr\"{o}dinger equation.
\begin{align}\label{eq:TimeDep_Sch_eq_app}
    i\hbar %\frac{\partial}{\partial t}
           \partial_t \ket{ \Psi_{{\bm x}_c, {\bm k}_c}} 
         = \Bigl( {\hat{H}}^{(0)} + {\hat V}(t) \Bigr) \ket{ \Psi_{{\bm x}_c, {\bm k}_c}}, 
\end{align}
where $ {\hat V}(t) = \sum_{\beta} \delta m_\beta(t) {\hat H}'_\beta$.
Substituting the wave-packet wavefunction, Eq.~(\ref{eq:WavePacket}), into Eq.~(\ref{eq:TimeDep_Sch_eq_app}), 
and multiplying the Bloch eigenstate $\bra{ \psi^{(0)}_{m\bm k'}(\bm m_0)}$ on both sides of the equation, 
we get the following differential equation, 
% \begin{widetext}
% \begin{align}\label{eq:C_diff_eq1}
%      \int_{\bm k} e^{i ({\bm k}-{\bm k}')\cdot{\hat{\bm x}}} 
%      \left(
%             i\hbar {\dot C}_{n\bm k} \braket{u^{(0)}_{m\bm k'}(\bm m_0)| u^{(0)}_{n\bm k}(\bm m_0) } 
%           + i\hbar \sum_{l \ne n} 
%                    {\dot C}_{l\bm k} \braket{u^{(0)}_{m\bm k'}(\bm m_0)| u^{(0)}_{n\bm k}(\bm m_0) }
%      \right) 
%       &= 
%      \int_{\bm k} e^{i ({\bm k}-{\bm k}')\cdot{\bm x}} 
%                   \left(
%                           C_{n} \langle u^{'(0)}_{m}|{\hat{\mathcal H}}_{0}|u^{(0)}_{n} \rangle 
%                         + \sum_{l \ne n} {\tilde C}_{l} 
%                           \langle u^{'(0)}_{m}|{\hat{\mathcal H}}_{0}|u^{(0)}_{l} \rangle
%                   \right) \\ \notag
%       & \ \ \ \ \ \ \ \ \ \ \ 
%     +\int_{\bm k} e^{i ({\bm k}-{\bm k}')\cdot{\bm x}} 
%                   \left(
%                           C_{n} \langle u^{'(0)}_{m}|{\hat{\mathcal H}}_{1}|u^{(0)}_{n} \rangle 
%                         + \sum_{l \ne n} {\tilde C}_{l} 
%                           \langle u^{'(0)}_{m}|{\hat{\mathcal H}}_{1}|u^{(0)}_{l} \rangle
%                   \right).
% \end{align}
% \end{widetext}
% 
% For the terms on the right-hand side of Eq.~(\ref{eq:Ctil_diff_eq1}) involving ${\hat{\mathcal H}}_{1}$, 
% we insert a complete set of $\sum_{n}|u^{(0)}_{n} \rangle \langle u^{(0)}_{n} |=1$.
% Then, by employing the orthogonal relation between Bloch state, $\langle u^{'(0)}_{m}| u^{(0)}_{n} \rangle = \delta({\bm k}-{\bm k}')\delta_{mn}$, the momentum integration can be carried out, and we find
%
\begin{widetext}
\begin{align}\label{eq:C_diff_eq}
      i\hbar \partial_t C_{n\bm k} \delta_{mn}
    + i\hbar \sum_{l \ne n}  \partial_t C_{l\bm k} \delta_{ml} 
      &=   
           \varepsilon^{(0)}_{n\bm k}(\bm x_c,\bm m_0) C_{n\bm k} \delta_{mn} 
         + \sum_{l (\ne n)} \varepsilon^{(0)}_{l\bm k}(\bm x_c,\bm m_0) C_{l\bm k} \delta_{ml} 
      % \\ \notag 
      % &\ \ \ \ \ 
         + ({\hat V}(t))_{ln} C_{n\bm k} 
         + \sum_{l (\ne n)} ({\hat V}(t))_{ml} C_{l\bm k}, 
\end{align}
\end{widetext}
%
% To this end, we omit the dependencies of the center of mass and time for the coefficient $C_{l\bm k}$ for notational simplicity.
Here, we use the orthogonal relation, 
$\braket{u^{(0)}_{m\bm k'}(\bm m_0)| u^{(0)}_{n\bm k}(\bm m_0) } = \delta(\bm k - \bm k')\delta_{mn}$.
One can separate Eq.(\ref{eq:C_diff_eq}) into the intra-band contribution and inter-band contribution ($m \ne n$), 
\begin{widetext}
\begin{align}\label{eq:C_diffEq_intra}
    i\hbar \partial_t C_{n\bm k} 
    &= 
     \varepsilon^{(0)}_{n\bm k}(\bm x_c,\bm m_0) C_{n\bm k} 
   + ({\hat V}(t))_{nn} C_{n\bm k} 
   + \sum_{l (\ne n)} ({\hat V}(t))_{nl} C_{l\bm k},
    \\
\label{eq:C_diffEq_inter}
    i\hbar \partial_t C_{m\bm k} 
    &= 
     \varepsilon^{(0)}_{m\bm k}(\bm x_c,\bm m_0) C_{m\bm k} 
   + ({\hat V}(t))_{mn} C_{n\bm k} 
   + \sum_{l (\ne n)} ({\hat V}(t))_{ml} C_{l\bm k} .
\end{align}
\end{widetext}
We need to solve the second differential equation, Eq.~(\ref{eq:C_diffEq_inter}).
To proceed, we assume that $C_{l{\bm k}}$ for $l\ne n$ differs from $C_{n{\bm k}}$ only by an overall factor $M_{ln,\bm k}$,
\begin{align}\label{eq:M_def_app}
    C_{l{\bm k}} 
      &= M_{ln,\bm k}~C_{n\bm k}.
\end{align}
Then, substituting Eq.(\ref{eq:M_def_app}) into Eq.(\ref{eq:C_diffEq_inter}), and after straightforward algebra, we arrive at
\begin{align}\label{eq:M_diffEq_app}
      i\hbar \partial_t M_{ln,\bm k} 
       &= 
        \varepsilon^{(0)}_{ln,\bm k}(\bm x_c,\bm m_0) M_{ln,\bm k} 
      + ( {\hat V}(t) )_{ln} 
      - ( {\hat V}(t) )_{nn} M_{ln,\bm k} \notag \\ 
       &
      + \sum_{m (\ne n)} ( {\hat V}(t) )_{lm} M_{mn,\bm k} 
      - \sum_{m (\ne n)} M_{ln,\bm k} ( {\hat V}(t) )_{nm} M_{mn,\bm k}, 
\end{align}

The coefficient $M_{ln,\bm k}$ is expanded in powers of the precession amplitude as $M_{ln,\bm k} = \sum_{N \ge 1} M^{(N)}_{ln,\bm k}$, and $M^{(N)}_{ln,\bm k} \propto \mathcal O (\delta m^{N})$. 
% 
% Substituting this series into Eq.~(\ref{eq:M_diffEq_app}) and collecting terms order by order. 
% We find that only the first-order coefficient is required for practical calculations of the wave-packet energy and the current (see following section), therefore focus on $M^{(1)}_{ln}$. 
Noting that ${\hat H}'_{\beta}$ is already ${\mathcal O} (\delta m^{1})$, the differential equation for $M^{(1)}_{ln,\bm k}$ reads, 
\begin{align}\label{eq:M1_diffEq_app}
      i\hbar \partial_t M^{(1)}_{ln,\bm k} 
      &= 
        \varepsilon^{(0)}_{ln,\bm k}(\bm x_c,\bm m_0) M^{(1)}_{ln,\bm k} 
      + ( {\hat V}(t) )_{ln}.
\end{align}
%
% 
% Employing the representation in Eq.~(\ref{eq:mag_vec}), one obtains the following solution to Eq.~(\ref{eq:M1_diffEq}), 
% %
% \begin{align}
%      M^{(1)}_{ln}({\bm k},t) &=  -J ~\frac{ \varepsilon^{(0)}_{ln} \delta m_{\beta}(t) - i \delta {\dot m}_{\beta}(t) }
%                                                  { \varepsilon^{(0)}_{ln} - \omega^{2}}  
%                                             \sigma^{ln}_{\beta}.
% \end{align}
% %
% 
% 
% 
Employing the Fourier transformations for the time-varying magnetization and $M^{(1)}_{ln,\bm k}$, 
\begin{align}\label{eq:Fourier_app}
      \delta \bm m(t) &= \sum_{\Omega}\delta \bm m_{\Omega} e^{-i\Omega t}, \\
      M^{(1)}_{ln,\bm k}(\bm x_c, \bm k_c, t) &= \sum_{\Omega}  M^{(1)}_{ln,\bm k}(\bm x_c, \bm k_c, \Omega) e^{-i\Omega t}
\end{align}
we find that the solution of Eq.(\ref{eq:M1_diffEq_app}) reads, 
\begin{align}\label{eq:M(1)_solution_app}
      M^{(1)}_{ln,\bm k}(\bm x_c, \bm k_c, t) 
      &= - \sum_{\Omega} \frac{ \delta m_{\Omega,\beta}~e^{-i\Omega t} }{ \varepsilon^{(0)}_{ln,\bm k}(\bm x_c,\bm m_0) - \hbar\Omega }( {\hat H}'_{\beta} )_{ln}.
\end{align}
% Then, considering the situation where the magnetization varies on time slowly compared with the energy scale of the electron system, we assume 
% $ \varepsilon^{(0)}_{ln,\bm k}(\bm x_c,\bm m_0) \gg \hbar\Omega$ and arrive at the following expression, 
% % 
% \begin{align}\label{eq:M(1)_solution2_app}
%       M^{(1)}_{ln,\bm k}(\bm x_c, \bm k_c, t) 
%       &= ( {\hat H}'_{\beta} )_{ln} \sum_{N=0}^{\infty} \left( \frac{i\hbar}{\varepsilon^{(0)}_{ln,\bm k}(\bm x_c,\bm m_0)} \right)^N 
%           \frac{\partial^{N}_t \delta m_{\beta}(t)}{\varepsilon^{(0)}_{ln,\bm k}(\bm x_c,\bm m_0)} 
% \end{align}
% 

\section{ Derivation of Eq.~(\ref{eq:Lagrangian}) }\label{sec:Appendix.B}

The Lagrangian is defined as 
\begin{align}\label{eq:Lagrangian_app}
     L &= \bra{\overline{ \Psi_{{\bm x}_c, {\bm k}_c}}} 
             \left( i\hbar\frac{d}{dt} - {\hat H}(t) \right) 
             \ket{\overline{ \Psi_{{\bm x}_c, {\bm k}_c}}}.
\end{align}
The time-dependencies are included in the coefficients $C_{n\bm k}$ of the wave-packet and the perturbation-included Bloch states $\ket{{\tilde u}_{n\bm k}(\bm x_c,\bm m)}$.
Hence, we can expand the total derivative with respect to time and separate the terms as follow,
\begin{widetext}
    \begin{align}\label{eq:Lagrangian_expand_app}
     L &= \bra{\overline{ \Psi_{{\bm x}_c, {\bm k}_c}}} 
             i\hbar \partial_t
          \ket{\overline{ \Psi_{{\bm x}_c, {\bm k}_c}}}
         + i\hbar{\dot {\bm m}} \cdot
          \bra{\overline{ \Psi_{{\bm x}_c, {\bm k}_c}}} 
             \partial_{\bm m}
          \ket{\overline{ \Psi_{{\bm x}_c, {\bm k}_c}}} 
         -\bra{\overline{ \Psi_{{\bm x}_c, {\bm k}_c}}} 
             \hat H(t)
          \ket{\overline{ \Psi_{{\bm x}_c, {\bm k}_c}}}  \notag \\
       &= \bra{ \Psi^{(0)}_{{\bm x}_c, {\bm k}_c}} 
             i\hbar \partial_t
          \ket{ \Psi^{(0)}_{{\bm x}_c, {\bm k}_c}}
         + i\hbar{\dot {\bm m}} \cdot
          \bra{\overline{ \Psi_{{\bm x}_c, {\bm k}_c}}} 
             \partial_{\bm m}
          \ket{\overline{ \Psi_{{\bm x}_c, {\bm k}_c}}} \notag \\
       &\ \ \ \ \   
         -\left(
                 \bra{ \Psi^{(0)}_{{\bm x}_c, {\bm k}_c}} 
                    i\hbar \partial_t
                 \ket{ \Psi^{(0)}_{{\bm x}_c, {\bm k}_c}} 
                -\bra{\overline{ \Psi_{{\bm x}_c, {\bm k}_c}}} 
                    i\hbar \partial_t
                 \ket{\overline{ \Psi_{{\bm x}_c, {\bm k}_c}}}
                +\bra{\overline{ \Psi_{{\bm x}_c, {\bm k}_c}}} 
             \hat H(t)
          \ket{\overline{ \Psi_{{\bm x}_c, {\bm k}_c}}}
          \right)
\end{align}
\end{widetext}
where we neglect the derivative with respect to $\bm x_c$ because, under the approximation that discards terms of order $(\delta m \bm \nabla_{\bm x_c})$ and higher, it is not essential for the present analysis.
The third to the fifth terms in Eq.(\ref{eq:Lagrangian_expand_app}) are the wave-packet energy ${\tilde {\mathcal E}}_n$ and evaluated in Appendix.\ref{sec:Appendix.C}.

The first term in Eq.(\ref{eq:Lagrangian_expand_app}) can read as follow by using the phase factor of the coefficient $C_{n\bm k}$; $C_{n\bm k} = |C_{n\bm k}|e^{-i\gamma_{n\bm k}}$, 
\begin{align}\label{eq:Lag_1st_term_app}
    &\bra{ \Psi^{(0)}_{{\bm x}_c, {\bm k}_c}} i\hbar \partial_t \ket{ \Psi^{(0)}_{{\bm x}_c, {\bm k}_c}} \notag \\
    &\ \ \ 
     = i\hbar \int_{\bm k}|C_{n\bm k}|\partial_t |C_{n\bm k}| 
      +\hbar \int_{\bm k} |C_{n\bm k}|^2 \partial_t \gamma_{n\bm k}
       \notag \\
    &\ \ \ 
     =
    \hbar \partial_t \gamma_{n\bm k_c} \notag \\
    &\ \ \ 
     = \hbar \frac{d}{dt}\gamma_{n\bm k_c} 
     - \hbar \dot{\bm k}_c \cdot \partial_{\bm k_c}\gamma_{n\bm k_c}
     - \hbar \dot{\bm x}_c \cdot \partial_{\bm x_c}\gamma_{n\bm k_c}
\end{align}
In the second line of Eq.(\ref{eq:Lag_1st_term_app}), we use the normalization condition for the coefficient; $\int_{\bm k}|C_{n\bm k}|^2 = 1 -\sum_{l(\ne n)}\int_{\bm k}|C_{l\bm k}|^2 \approx 1$.
% 
% \red{Is it necessary to require that the coefficient of the $n$-th band; $C_{n\bm k}$ explicitly depends on $\bm x_c$?}\blue{
% The third term in the last line of Eq.(\ref{eq:Lag_1st_term_app}) is proportional to second or higher order in the product of $\delta m$ and the spatial gradient, when substituting a derived $\bm x_c$'s equation of motion.
% % 
% In the following, we ignore this term, since it is not main concern in our work.
% 
In the following, we ignore the third term in the last line of Eq.(\ref{eq:Lag_1st_term_app}) by assuming the approximation described above.
The second term of the third line in Eq.(\ref{eq:Lag_1st_term_app}) can be rewritten by using the condition ${\bm x}_c = \bra{\Psi_{{\bm x}_c, {\bm k}_c}} \hat {\bm x} \ket{\Psi_{{\bm x}_c, {\bm k}_c}}$ as follows.
Substituting the wave-packet wavefunction and using the expression of the matrix elements of the position operator \footnote{E. I. Blount, in Solid State Physics, edited by F. Seitz and D.
Turnbull Academic Press, New York, 1962 , Vol. 13, p. 305.} 
\begin{align}
     \langle \psi^{(0)}_{n{\bm q}} |{\hat {\bm x}}| \psi^{(0)}_{n'{\bm q}'} \rangle 
      &=
     \left\{ i\frac{\partial }{\partial {\bm q}}\delta_{n n'} + \langle u^{(0)}_{n\bm q}| i\partial_{\bm q} |u^{(0)}_{n'\bm q} \rangle \right\}\delta({\bm q}-{\bm q}'),
\end{align}
we find that this condition can read 
% 同じような計算をしている論文を引用する．Sundaram & Niu, Jinxiong Jia (2024)など
% 
\begin{align}
    \bm x_c &= \partial_{\bm k_c}\gamma_{n\bm k_c} -{\tilde{\bm A}}^{\bm k_c }_{n},
              % - {\bm A}^{\bm k_c (0)}_{n} 
              % - {\bm A}^{\bm k_c (1)}_{n} 
              % - {\bm A}^{\bm k_c (2)}_{n}
\end{align}
where the perturbation-included Berry connections in the momentum space are introduced as 
${\tilde{\bm A}}^{\bm k_c }_{n} = -i \bra{{\tilde u}_{n\bm k_c} (\bm x_c,\bm m)} \partial_{\bm k_c}\ket{{\tilde u}_{n\bm k_c} (\bm x_c,\bm m)}  $
% = {\bm A}^{\bm k_c (0)}_{n} +{\bm A}^{\bm k_c (1)}_{n} + {\bm A}^{\bm k_c (2)}_{n} + \cdots$. 
% and each term is defined as 
% \begin{align}
%     {\bm A}^{\bm k_c (1)}_{n} &= 2{\rm Re} 
%                                 \left[ 
%                                      -i \bra{u^{(1)}_{n\bm k_c} (\bm m)} 
%                                         \partial_{\bm k_c}
%                                         \ket{u^{(0)}_{n\bm k_c} (\bm m_0)}
%                                 \right], \\
%     {\bm A}^{\bm k_c (2)}_{n} &= 2{\rm Re} 
%                                 \left[ 
%                                      -i \bra{u^{(2)}_{n\bm k_c} (\bm m)} 
%                                         \partial_{\bm k_c}
%                                         \ket{u^{(0)}_{n\bm k_c} (\bm m_0)}
%                                 \right]   \notag \\
%                               &\ \ \ \ \ \ \ \ \ \ \ 
%                                  -i \bra{u^{(1)}_{n\bm k_c} (\bm m)} 
%                                         \partial_{\bm k_c}
%                                         \ket{u^{(1)}_{n\bm k_c} (\bm m)}    \notag \\
%                               &\ \ \ \ \ \ \ \ \ \ \ 
%                                  -i \delta\bra{u^{(0)}_{n\bm k_c} (\bm m_0)} 
%                                         \partial_{\bm k_c}
%                                         \ket{u^{(0)}_{n\bm k_c} (\bm m_0)}.
% \end{align}
Hence Eq.(\ref{eq:Lag_1st_term_app}) eventually reads 
% \begin{widetext}
%     \begin{align}
%     \bra{ \Psi^{(0)}_{{\bm x}_c, {\bm k}_c}} i\hbar \partial_t \ket{ \Psi^{(0)}_{{\bm x}_c, {\bm k}_c}} 
% % 
%     &= \hbar \frac{d}{dt}\gamma_{n\bm k_c}(\bm x_c, \bm k_c, t)  
%      - \hbar \dot{\bm k}_c \cdot 
%        \left( 
%              \bm x_c + {\bm A}^{\bm k_c (0)}_{n} + {\bm A}^{\bm k_c (1)}_{n} + {\bm A}^{\bm k_c (2)}_{n}
%        \right) \notag \\
% % 
%     &=\hbar \frac{d}{dt} \left( \gamma_{n\bm k_c}(\bm x_c, \bm k_c, t) -\bm x_c \cdot \bm k_c  \right)
%      +\hbar \dot{\bm x}_c \cdot \bm k_c
%      -\hbar \dot{\bm k}_c \cdot 
%        \left( 
%               {\bm A}^{\bm k_c (0)}_{n} + {\bm A}^{\bm k_c (1)}_{n} + {\bm A}^{\bm k_c (2)}_{n}
%        \right)
% \end{align}
% \end{widetext}
% 
\begin{align}\label{eq:Lagran_1st_term2_app}
    &\bra{ \Psi^{(0)}_{{\bm x}_c, {\bm k}_c}} i\hbar \partial_t \ket{ \Psi^{(0)}_{{\bm x}_c, {\bm k}_c}} \notag \\
    &\ \ \ \ 
     = \hbar \frac{d}{dt}\gamma_{n\bm k_c}  
     - \hbar \dot{\bm k}_c \cdot 
       \left( 
             \bm x_c + {\tilde{\bm A}}^{\bm k_c }_{n}
       \right) \notag \\
    &\ \ \ \ 
     =\hbar \frac{d}{dt} \left( \gamma_{n\bm k_c} -\bm x_c \cdot \bm k_c  \right)
     +\hbar \dot{\bm x}_c \cdot \bm k_c
     -\hbar \dot{\bm k}_c \cdot {\tilde{\bm A}}^{\bm k_c }_{n}.
\end{align}

On the other hand, the second term in Eq.(\ref{eq:Lagrangian_expand_app}) is expressed by introducing the perturbation-included Berry connection in the magnetization space 
${\tilde{\bm A}}^{\bm m}_{n} = -i \bra{{\tilde u}_{n\bm k_c} (\bm x_c,\bm m)} \partial_{\bm m}\ket{{\tilde u}_{n\bm k_c} (\bm x_c,\bm m)}  $,
% \begin{widetext}
% \begin{align}
%     i\hbar{\dot m}_{\alpha}
%           \bra{\overline{ \Psi_{{\bm x}_c, {\bm k}_c}}} 
%              \partial_{m_\alpha}
%           \ket{\overline{ \Psi_{{\bm x}_c, {\bm k}_c}}}
%     &= i\hbar{\dot m}_{\alpha} 
%        \int_{\bm k} \int_{\bm k'} C^*_{n\bm k} C_{n\bm k'}
%        \left( 
%              \bra{{\tilde u}_{n\bm k_c} (\bm m)} + \delta \bra{u^{(0)}_{n\bm k_c} (\bm m_0)} 
%        \right)
%        \partial_{m_\alpha}
%        \left(
%              \ket{{\tilde u}_{n\bm k_c} (\bm m)} + \delta \ket{u^{(0)}_{n\bm k_c} (\bm m_0)} 
%        \right)  \notag \\
% % 
%     &= -\hbar {\dot m}_{\alpha} {\tilde A}^{\bm m}_{n,\alpha}
% \end{align}
% \end{widetext}
\begin{align}\label{eq:Lagran_2nd_term_app}
    i\hbar{\dot {\bm m}} \cdot
          \bra{\overline{ \Psi_{{\bm x}_c, {\bm k}_c}}} 
             \partial_{\bm m}
          \ket{\overline{ \Psi_{{\bm x}_c, {\bm k}_c}}}
    &= -\hbar {\dot {\bm m}} \cdot {\tilde {\bm A}}^{\bm m}_n
\end{align}
If the electrostatic potential contained in the static Hamiltonian is written explicitly as a separate term, 
then by combining Eqs. (\ref{eq:Lagran_1st_term2_app}) and (\ref{eq:Lagran_2nd_term_app}), the Lagrangian can finally be written as the form of Eq.(\ref{eq:Lagrangian}) in the main text.

\section{ Derivation of Eq.~(\ref{eq:WavePac_En(2)})} \label{sec:Appendix.C}

Here we derive the wave-packet energy up to the order of $\delta m^2$. 
Its definition is as follow,
\begin{align}
    &\label{eq:En_WP+dyn_app}
    {\tilde{\mathcal E}}_n = {\tilde{\mathcal E}}^{\rm WP}_n + {\tilde{\mathcal E}}^{\rm dyn}_n, \\
    &\label{eq:En_WP_app}
    {\tilde{\mathcal E}}^{\rm WP}_n = \bra{ \overline{ \Psi_{{\bm x}_c, {\bm k}_c}} }
                                        \hat H(t) 
                                       \ket{ \overline{ \Psi_{{\bm x}_c, {\bm k}_c}} }, \\
    &\label{eq:En_dyn_app}
    {\tilde{\mathcal E}}^{\rm dyn}_n = 
              \bra{ \Psi^{(0)}_{{\bm x}_c, {\bm k}_c} } 
                   i\hbar\partial_t 
              \ket{ \Psi^{(0)}_{{\bm x}_c, {\bm k}_c} }
            - \bra{ \overline{ \Psi_{{\bm x}_c, {\bm k}_c}} }
                   i\hbar \partial_t 
              \ket{ \overline{ \Psi_{{\bm x}_c, {\bm k}_c}} }.
\end{align}

Substituting the Hamiltonian and the wave-packet wavefunction into Eq.(\ref{eq:En_WP_app}) and summing the terms up to the second-order of perturbation, 
we find the following expression,
\begin{align}\label{eq:EnWP_1_app}
    {\mathcal E}^{{\rm WP}(2)}_n 
     &= 
     2\delta \bra{ \Psi^{(0)}_{{\bm x}_c, {\bm k}_c} } {\hat H}^{(0)} \ket{ \Psi^{(0)}_{{\bm x}_c, {\bm k}_c} } 
    + \bra{ \Psi^{(1)}_{{\bm x}_c, {\bm k}_c} } {\hat H}^{(0)} \ket{ \Psi^{(1)}_{{\bm x}_c, {\bm k}_c} } \notag \\
     &\ \ \ \ \ 
    + \sum_\beta2{\rm Re}\left[ \bra{ \Psi^{(0)}_{{\bm x}_c, {\bm k}_c} }  \delta m_\beta (t){\hat H}'_{\beta} \ket{ \Psi^{(1)}_{{\bm x}_c, {\bm k}_c} } \right]
\end{align}
The first line in Eq.(\ref{eq:EnWP_1_app}) can rewritten as 
\begin{align}
    &2\delta \bra{ \Psi^{(0)}_{{\bm x}_c, {\bm k}_c} } {\hat H}^{(0)} \ket{ \Psi^{(0)}_{{\bm x}_c, {\bm k}_c} } 
    + \bra{ \Psi^{(1)}_{{\bm x}_c, {\bm k}_c} } {\hat H}^{(0)} \ket{ \Psi^{(1)}_{{\bm x}_c, {\bm k}_c} } \notag \\
    &\ \ \ \ =
     \sum_{l(\ne n)} \varepsilon^{(0)}_{ln}(\bm x_c, \bm m_0) M^{(1)\ast}_{ln,\bm k_c}M^{(1)}_{ln,\bm k_c},
\end{align}
while the second line in Eq.(\ref{eq:EnWP_1_app}) can be rewritten by substituting the wave-packet wavefunction 
\begin{align}\label{eq:EnWP_secondline_app}
    &\sum_\beta2{\rm Re}\left[ \bra{ \Psi^{(0)}_{{\bm x}_c, {\bm k}_c} } \delta m_\beta (t){\hat H}'_{\beta} \ket{ \Psi^{(1)}_{{\bm x}_c, {\bm k}_c} } \right] \notag \\
    & =
     \sum_\beta2{\rm Re}\left[ 
                   \int_{\bm k}\int_{\bm k'} C^*_{n\bm k}\sum_{l(\ne n)}M^{(1)}_{ln,\bm k'}C_{n\bm k'} 
                   \right.\notag \\
    &\left. \ \ \ \ \ \ \ \ \ \ \ \ \ \ \ \ \ \ \ \ \ \ \ \times
                   \bra{\psi^{(0)}_{n\bm k}(\bm m_0)}
                       \delta m_\beta (t){\hat H}'_{\beta}
                   \ket{\psi^{(0)}_{l\bm k'}(\bm m_0)}
              \right] \notag \\
    & = 
     \sum_\beta2{\rm Re}\left[ 
                   \sum_{l(\ne n)}M^{(1)}_{ln,\bm k_c} ({\hat H}'_{\beta})_{ln}
              \right]\delta m_\beta (t).
\end{align}

On the other hand, the dynamical energy is calculated up to the second-order as follow, 
\begin{align}\label{eq:EnDyn_app}
    {\mathcal E}^{{\rm dyn}(2)}_n &= 
    -\delta \bra{ \Psi^{(0)}_{{\bm x}_c, {\bm k}_c} } 
                i\hbar\partial_t 
            \ket{ \Psi^{(0)}_{{\bm x}_c, {\bm k}_c} }
    - \bra{ \Psi^{(0)}_{{\bm x}_c, {\bm k}_c} } 
               ( i\hbar\partial_t \delta )
            \ket{ \Psi^{(0)}_{{\bm x}_c, {\bm k}_c} } \notag \\
    &\ \ \ \ 
    - \bra{ \Psi^{(1)}_{{\bm x}_c, {\bm k}_c} } 
                i\hbar\partial_t 
      \ket{ \Psi^{(1)}_{{\bm x}_c, {\bm k}_c} }.
\end{align}
The first line in Eq.(\ref{eq:EnDyn_app}) can be expressed as follow by substituting the wave-packet wavefunction,
\begin{align}\label{eq:EnDyn_1stline_app}
    &-\delta \bra{ \Psi^{(0)}_{{\bm x}_c, {\bm k}_c} } 
                i\hbar\partial_t 
            \ket{ \Psi^{(0)}_{{\bm x}_c, {\bm k}_c} }
    - \bra{ \Psi^{(0)}_{{\bm x}_c, {\bm k}_c} } 
               ( i\hbar\partial_t \delta )
            \ket{ \Psi^{(0)}_{{\bm x}_c, {\bm k}_c} } \notag \\
    &\ \ \ \ \ \ 
    =  -2\hbar\delta\partial_t \gamma_{n\bm k_c} - i\hbar \partial_t \delta.
\end{align}
The second line is somewhat complicated but can be expressed by some algebras, 
\begin{align}\label{eq:EnDyn_2ndline_app}
    &\bra{ \Psi^{(1)}_{{\bm x}_c, {\bm k}_c} } 
                i\hbar\partial_t 
      \ket{ \Psi^{(1)}_{{\bm x}_c, {\bm k}_c} }  \notag \\
    &\ \ \ 
    = \frac{1}{2} \sum_{l(\ne n)} 
      \left(
             M^{(1)\ast}_{ln,\bm k_c} (i\hbar\partial_{t} M^{(1)}_{ln,\bm k_c} )
            -(i\hbar\partial_{t} M^{(1)\ast}_{ln,\bm k_c})  M^{(1)}_{ln,\bm k_c}  
      \right) \notag \\
    &\ \ \ \ \ \ 
       - i\hbar\partial_{t}\delta 
       -2\hbar \delta \partial_t \gamma_{n\bm k_c}.
\end{align}                      
Hence the dynamical energy can be summarized by accompanying Eqs.(\ref{eq:EnDyn_1stline_app}) and (\ref{eq:EnDyn_2ndline_app}) as, 
\begin{align}\label{eq:EnDyn_2_app}
    {\mathcal E}^{{\rm dyn}(2)}_n &= 
    -\frac{1}{2} \sum_{l(\ne n)} 
      \left(
             M^{(1)\ast}_{ln,\bm k_c} (i\hbar\partial_{t} M^{(1)}_{ln,\bm k_c} )
            -(i\hbar\partial_{t} M^{(1)\ast}_{ln,\bm k_c})  M^{(1)}_{ln,\bm k_c}  
      \right).
\end{align}

If the Hamiltonian carries no $\bm x_c$-dependence and the magnetization precesses around one axis, then the explicit form of the perturbation coefficient $M^{(1)}_{ln,\bm k_c}$ is known. 
Substituting it into Eqs.(\ref{eq:EnWP_secondline_app}) and (\ref{eq:EnDyn_2_app}), the energy of the wave-packet can be summarized as follows.
\begin{widetext}
\begin{align}
    { {\mathcal E}}^{(2)}_n  
     &= 
     \sum_{l(\ne n)} \varepsilon^{(0)}_{ln}(\bm x_c, \bm m_0) M^{(1)\ast}_{ln,\bm k_c}M^{(1)}_{ln,\bm k_c}
    + \sum_\beta2{\rm Re}\left[ 
                   \sum_{l(\ne n)}M^{(1)}_{ln,\bm k_c} ({\hat H}'_{\beta})_{ln}
              \right]\delta m_\beta (t)
    -\frac{1}{2} \sum_{l(\ne n)} 
      \left(
             M^{(1)\ast}_{ln,\bm k} (i\hbar\partial_{t} M^{(1)}_{ln,\bm k} )
            -(i\hbar\partial_{t} M^{(1)\ast}_{ln,\bm k})  M^{(1)}_{ln,\bm k}  
      \right) \notag \\
     &=  J_{\rm ex}^{2}\sum_{\alpha, \beta}{\rm Re} \left[
                                               \sum_{l(\ne n)}  
                                               \frac{ \varepsilon^{(0)}_{nl}(\bm x_c, \bm m_0) \sigma^{nl}_{\alpha} \sigma^{ln}_{\beta}}
                                                    {(\varepsilon^{(0)}_{nl}(\bm x_c, \bm m_0))^2 - \hbar^2\omega^{2}}
                                         \right] \delta m_{\alpha}(t)\delta m_{\beta}(t) 
                               +J_{\rm ex}^{2}\sum_{\alpha,\beta}{\rm Im} \left[
                                               \sum_{l(\ne n)} 
                                               \frac{ \sigma^{nl}_{\alpha} \sigma^{ln}_{\beta} }
                                                    { (\varepsilon^{(0)}_{nl}(\bm x_c, \bm m_0))^2 - \hbar^2\omega^{2}}
                                         \right] \delta m_{\alpha}(t)\delta {\dot m}_{\beta}(t).
\end{align}
\end{widetext}

\section{Definition of normalized quantum geometric tensor}\label{sec:Appendix.DefOfQGT}

Since the normalized quantum geometric quantities are used both in the main text and in this Appendix, 
we summarize them here.
To this end, we consider the situation as the same as Sec. \ref{sec:NLcurrent} in the main text, where $\bm x_c$-dependence no longer exists.
The conventional Quantum geometric tensor is defined as 
${\mathcal Q}^{\bm \xi \bm \xi', nl}_{\mu \nu} = {\mathcal A}^{\bm \xi, nl}_{\mu}{\mathcal A}^{\bm \xi', ln}_{\nu}$, where ${\mathcal A}^{\bm \xi, nl}_{\mu} = -i \bra{u^{(0)}_{n\bm k}(\bm m_0)}\partial_{\xi_\mu}\ket{u^{(0)}_{n\bm k}(\bm m_0)}$.
On the other hand, the ``normalized'' quantum geometric tensor is defined as
\begin{align}
    {\overline{\mathcal Q}}^{\bm \xi \bm \xi', nl}_{\mu \nu} 
    &=\frac{( \varepsilon^{(0)}_{nl,\bm k}(\bm m_0) )^2}{ ( \varepsilon^{(0)}_{nl,\bm k}(\bm m_0) )^2 - \hbar^2 \omega^2}
      {\mathcal A}^{\bm \xi, nl}_{\mu}{\mathcal A}^{\bm \xi', ln}_{\nu}.
\end{align}
As same as the conventional quantum geometric tensor, the real (imaginary) part of ${\overline{\mathcal Q}}^{\bm \xi \bm \xi', nl}_{\mu \nu}$ corresponds to the quantum metric (Berry curvature), 
\begin{align}
    {\overline {g}}^{\bm \xi\bm \xi'}_{n,\mu\nu} 
      &= {\rm Re} \left[ \sum_{l(\ne n)} {\overline{\mathcal Q}}^{\bm \xi  , nl}_{\mu \nu} \right] \notag \\
      &= {\rm Re} \left[
                       \sum_{l (\ne n)} 
                       \frac{( \varepsilon^{(0)}_{nl,\bm k}(\bm m_0) )^2{\mathcal A}^{{\bm \xi},nl}_{\mu} {\mathcal A}^{{\bm \xi'},ln}_{\nu} }
                            { ( \varepsilon^{(0)}_{nl,\bm k}(\bm m_0) )^2 - \hbar^2\omega^{2} }
                  \right], \\
    {\overline {\Omega}}^{\bm \xi \bm \xi'}_{n,\mu\nu} 
     &= -2{\rm Im}\left[\sum_{l(\ne n)}  {\overline{\mathcal Q}}^{\bm \xi \bm \xi', nl}_{\mu \nu}\right] \notag \\
     &= -2{\rm Im} \left[
                       \sum_{l (\ne n)} 
                       \frac{( \varepsilon^{(0)}_{nl,\bm k}(\bm m_0) )^2{\mathcal A}^{{\bm \xi},nl}_{\mu} {\mathcal A}^{{\bm \xi'},ln}_{\nu} }
                            { ( \varepsilon^{(0)}_{nl,\bm k}(\bm m_0) )^2 - \hbar^2\omega^{2} }.
                  \right]
\end{align}
The Berry connection polarizability is defined as  
\begin{align}
    {\overline G}^{\bm \xi\bm \xi'}_{n,\mu\nu} 
    &= {\rm Re} \left[
                       \sum_{l (\ne n)} 
                       \frac{\varepsilon^{(0)}_{nl}(\bm m_0){\mathcal A}^{{\bm \xi},nl}_{\mu} {\mathcal A}^{{\bm \xi'},ln}_{\nu} }
                            { ( \varepsilon^{(0)}_{nl,\bm k}(\bm m_0) )^2 - \hbar^2\omega^{2} }
                  \right].
\end{align}

\section{Evaluation of first order correction for Berry connection}\label{sec:Appendix.A(1)derive}

To compute the current based on Eq.~(\ref{eq:wavepac_x_EOM}), 
the first-order correction in the perturbation to ${\tilde \Omega}^{{\bm k}{\bm m}}_{n,i\alpha}$ is required.
As defined in Eq.~(\ref{eq:BC_def}), 
${\tilde \Omega}^{{\bm k}{\bm m}}_{n,i\alpha}$ is constructed from the Berry connection including the perturbation. 
We introduce the first-order correction to the Berry curvature as follows.
\begin{align}\label{eq:BC_1stPerturb_app}
     \Omega^{{\bm k}{\bm m}(1)}_{n,i\alpha} &= \partial_{k_{i}} A^{{\bm m}(1)}_{n,\alpha} - \partial_{m_{\alpha}} A^{{\bm k}(1)}_{n,i}, 
\end{align}
where the first-order correction to the Berry connection for $\bm \xi = \bm k$ or $\bm m$ is introduced as follow, 
\begin{align}\label{eq:A(1)xi_1st_term_app}
     {\tilde A}^{{\bm \xi}}_{n,\mu} &= -i \langle {\tilde u}_{n\bm k}(\bm m) | \partial_{\xi_{\mu}} | {\tilde u}_{n\bm k} (\bm m)\rangle \notag \\
                          &= -i \langle u^{(0)}_{n\bm k}(\bm m_0) | \partial_{\xi_{\mu}} | u^{(0)}_{n\bm k}(\bm m_0) \rangle
                             -i \langle u^{(0)}_{n\bm k}(\bm m_0) | \partial_{\xi_{\mu}} | u^{(1)}_{n\bm k}(\bm m) \rangle \notag \\
                          &\ \ \ \ \ \ \ \ 
                            -i \langle u^{(1)}_{n\bm k}(\bm m) | \partial_{\xi_{\mu}} | u^{(0)}_{n\bm k}(\bm m_0) \rangle + \mathcal{O}(\delta m^{2}) \notag \\
                          &\cong
                             A^{{\bm \xi}(0)}_{\mu} + A^{{\bm \xi}(1)}_{\mu}, \\
     A^{{\bm \xi}(1)}_{n,\mu} &= -i \langle u^{(0)}_{n\bm k}(\bm m_0) | \partial_{\xi_{\mu}} | u^{(1)}_{n\bm k}(\bm m) \rangle 
                                -i \langle u^{(1)}_{n\bm k}(\bm m) | \partial_{\xi_{\mu}} | u^{(0)}_{n\bm k}(\bm m_0) \rangle \notag \\
                            &= 2{\rm Re}\left[ -i \langle u^{(1)}_{n\bm k}(\bm m) | \partial_{\xi_{\mu}} | u^{(0)}_{n\bm k}(\bm m_0) \rangle \right].   
\end{align}
In Eq.~(\ref{eq:A(1)xi_1st_term}), 
we first derive the first-order correction to the $\bm m$-space Berry connection.
By making use of Eq.~(\ref{eq:Bloch_pertub}), the calculation proceeds as follows
\begin{widetext}    
\begin{align}\label{eq:Am_(1)_app}
     A^{{\bm m}(1)}_{n,\alpha} &= \sum_\beta2 {\rm Re} \left[ 
                                                  \sum_{l (\ne n)} \frac{-i J_{\rm ex}^{2}\sigma^{nl}_{\alpha} \sigma^{ln}_{\beta}}{ ( \varepsilon^{(0)}_{nl,\bm k}(\bm m_0) )^2 - \hbar^2\omega^{2} }
                                                     \left(
                                                             \delta m_{\beta}(t)
                                                           -i\hbar\frac{ \delta {\dot m}_{\beta}(t) }{ \varepsilon^{(0)}_{nl,\bm k}(\bm m_0) }
                                                     \right)
                                          \right] \notag \\
                             &= \sum_\beta2 {\rm Im} \left[
                                                 \sum_{l (\ne n)} \frac{( \varepsilon^{(0)}_{nl,\bm k}(\bm m_0) )^2}{ ( \varepsilon^{(0)}_{nl,\bm k}(\bm m_0) )^2 - \hbar^2\omega^{2} }
                                                 {\mathcal A}^{{\bm m},nl}_{\alpha} {\mathcal A}^{{\bm m},ln}_{\beta}
                                          \right] \delta m_{\beta}(t) 
                            -2\hbar \sum_\beta{\rm Re} \left[
                                                 \sum_{l (\ne n)} \frac{( \varepsilon^{(0)}_{nl,\bm k}(\bm m_0) )^2}{ ( \varepsilon^{(0)}_{nl,\bm k}(\bm m_0) )^2 - \hbar^2\omega^{2} }
                                                 \frac{{\mathcal A}^{{\bm m},nl}_{\alpha} {\mathcal A}^{{\bm m},ln}_{\beta}}
                                                 {\varepsilon^{(0)}_{nl,\bm k}(\bm m_0) }
                                          \right] \delta {\dot m}_{\beta}(t)    \notag \\
                             &= - \sum_\beta {\bar \Omega}^{{\bm m}{\bm m}}_{n,\alpha \beta}\; \delta m_{\beta}(t)
                                -2 \hbar\sum_\beta {\bar G}^{{\bm m}{\bm m}}_{n,\alpha \beta}\; \delta {\dot m}_{\beta}(t)
\end{align}
\end{widetext}
Here, the relation 
\begin{align}\label{eq:sigma_Am_app}
    \sigma^{nl}_{\alpha} = \frac{\varepsilon^{(0)}_{nl,\bm k}(\bm m_0)}{J_{\rm ex}} i {\mathcal A}^{{\bm m},nl}_{\alpha}
\end{align}
where 
${\mathcal A}^{{\bm m},nl}_{\alpha} = -i \langle u^{(0)}_{n\bm k}(\bm m_0) | \partial_{m_{\alpha}} | u^{(0)}_{l\bm k}(\bm m_0) \rangle$
has been employed to derive the second line of Eq.~(\ref{eq:Am_(1)_app}).
% 
% 
% 
% \textcolor{blue}{2025/9/15}
% 
For the $\bm k$-space Berry connection, we calculate its first-order correction for the perturbation as well as the $\bm m$-space one.
Using the relation 
\begin{align}\label{eq:v_Ak_app}
    v^{nl}_{i} = -\varepsilon^{(0)}_{nl,\bm k}(\bm m_0) i {\mathcal A}^{{\bm k},nl}_{i}
\end{align} 
where ${\mathcal A}^{{\bm k},nl}_{i} = -i \langle u^{(0)}_{n\bm k}(\bm m_0) | \partial_{k_{i}} | u^{(0)}_{l\bm k}(\bm m_0) \rangle$, we get the following result, 
\begin{widetext}    
\begin{align}\label{eq:Ak_(1)_app}
     A^{{\bm k}(1)}_{n,j}      &= \sum_\beta2 {\rm Im} \left[
                                                 \sum_{l (\ne n)} \frac{( \varepsilon^{(0)}_{nl,\bm k}(\bm m_0) )^2}{ ( \varepsilon^{(0)}_{nl,\bm k}(\bm m_0) )^2 - \hbar^2\omega^{2} }
                                                 {\mathcal A}^{{\bm k},nl}_{j} {\mathcal A}^{{\bm m},ln}_{\beta}
                                          \right] \delta m_{\beta}(t) 
                               -2\sum_\beta {\rm Re} \left[
                                                 \sum_{l (\ne n)} \frac{( \varepsilon^{(0)}_{nl,\bm k}(\bm m_0) )^2}{ ( \varepsilon^{(0)}_{nl,\bm k}(\bm m_0) )^2 - \hbar^2\omega^{2} }
                                                 \frac{{\mathcal A}^{{\bm k},nl}_{j} {\mathcal A}^{{\bm m},ln}_{\beta}}
                                                 {\varepsilon^{(0)}_{nl,\bm k}(\bm m_0) }
                                          \right] \delta {\dot m}_{\beta}(t), \notag \\
                            &= -  \sum_\beta{\bar \Omega}^{{\bm k}{\bm m}}_{n,j \beta}\; \delta m_{\beta}(t)
                                -2 \sum_\beta{\bar G}^{{\bm k}{\bm m}}_{n,j \beta}\; \delta {\dot m}_{\beta}(t).
\end{align}
\end{widetext}

\begin{widetext}
\section{Formulae of precession current for numerical calculations}\label{sec:Appendix.I}

Derivatives of the quantum geometric quantities appear in the expressions for the nonlinear current. 
In the numerical calculations presented in Sec.\ref{seubsec:numerical}, these derivatives are rewritten in terms of matrix elements of the velocity or spin operators. 
By using Eqs.(\ref{eq:sigma_Am_app}) and (\ref{eq:v_Ak_app}), and the relation of the complete system with respect to $ \ket{ u^{(0)}_{n\bm k}(\bm m_0)} $, we get the following formulae, 

\begin{align}\label{eq:veloMat_diffM}
     \partial_{m_\alpha} v^{n l}_{i}
        &=  \bra{\partial_{m_\alpha} u^{(0)}_{n\bm k}(\bm m_0)} {\hat v}_{i} \ket{ u^{(0)}_{l\bm k}(\bm m_0)}
           +\bra{ u^{(0)}_{n\bm k}(\bm m_0)} {\hat v}_{i} \ket{\partial_{m_\alpha} u^{(0)}_{l\bm k}(\bm m_0)}
            \notag \\
        &= \sum_{m} \langle \partial_{m_\alpha} u^{(0)}_{n\bm k}(\bm m_0)| u^{(0)}_{m\bm k}(\bm m_0) \rangle 
                    \langle u^{(0)}_{m\bm k}(\bm m_0) | {\hat v}_{i} | u^{(0)}_{l\bm k}(\bm m_0) \rangle %\notag\\
%        &\ \ \ 
          +\sum_{m} \langle u^{(0)}_{n\bm k}(\bm m_0) | {\hat v}_{i} | u^{(0)}_{m\bm k}(\bm m_0) \rangle  
                    \langle u^{(0)}_{m\bm k}(\bm m_0) | \partial_{m_\alpha} u^{(0)}_{l\bm k}(\bm m_0)\rangle \notag\\
        &= -J_{\rm ex}\sum_{m(\ne n)}
            \frac{ \sigma^{n m}_{\alpha} v^{m l}_{i} }{ \varepsilon^{(0)}_{nm,\bm k}(\bm m_0) } 
          ~+J_{\rm ex}\sum_{m(\ne l)}
            \frac{ v^{n m}_{i}\sigma^{m l}_{\alpha} }{ \varepsilon^{(0)}_{ml,\bm k}(\bm m_0) } 
\end{align}
\begin{align}\label{eq:spinMat_diffK}
     \partial_{k_i} \sigma^{n l}_{\alpha}
        &=  \langle \partial_{k_i} u^{(0)}_{n\bm k}(\bm m_0) | {\hat\sigma}_{\alpha} | u^{(0)}_{l\bm k}(\bm m_0) \rangle
          + \langle u^{(0)}_{n\bm k}(\bm m_0) | {\hat\sigma}_{\alpha} | \partial_{k_i} u^{(0)}_{l\bm k}(\bm m_0)  \rangle \notag \\
        &= \sum_{m} \langle \partial_{k_i} u^{(0)}_{n\bm k}(\bm m_0) | u^{(0)}_{m\bm k}(\bm m_0) \rangle 
                    \langle u^{(0)}_{m\bm k}(\bm m_0) | {\hat\sigma}_{\alpha} | u^{(0)}_{l\bm k}(\bm m_0) \rangle
%        \notag \\
%        &\ \ \ 
          +\sum_{m} \langle u^{(0)}_{n\bm k}(\bm m_0) | {\hat\sigma}_{\alpha}| u^{(0)}_{m\bm k}(\bm m_0) \rangle  
                    \langle u^{(0)}_{m\bm k}(\bm m_0) | \partial_{k_i} u^{(0)}_{l\bm k}(\bm m_0)  \rangle \notag \\
        &= \sum_{m(\ne n)}  \frac{ v^{n m}_{i} \sigma^{m l}_{\alpha} }{ \varepsilon^{(0)}_{nm,\bm k}(\bm m_0) } 
          -\sum_{m(\ne l)}  \frac{ \sigma^{n m}_{\alpha} v^{m l}_{i} }{ \varepsilon^{(0)}_{ml,\bm k}(\bm m_0) } 
\end{align}
\begin{align}\label{eq:spinMat_diffM}
     \partial_{m_\alpha} \sigma^{l n}_{\beta}
        &=  \langle \partial_{m_\alpha} u^{(0)}_{l\bm k}(\bm m_0) | {\hat\sigma}_{\beta} | u^{(0)}_{n\bm k}(\bm m_0) \rangle
           +\langle u^{(0)}_{l\bm k}(\bm m_0) | {\hat\sigma}_{\beta} | \partial_{m_\alpha} u^{(0)}_{n\bm k}(\bm m_0) \rangle \notag \\
        &= \sum_{m} \langle \partial_{m_\alpha} u^{(0)}_{l\bm k}(\bm m_0) | u^{(0)}_{m\bm k}(\bm m_0) \rangle 
                    \langle u^{(0)}_{m\bm k}(\bm m_0) | {\hat\sigma}_{\beta} | u^{(0)}_{n\bm k}(\bm m_0) \rangle
%        \notag \\
%        &\ \ \ 
          +\sum_{m} \langle u^{(0)}_{l\bm k}(\bm m_0) | {\hat\sigma}_{\beta} | u^{(0)}_{m\bm k}(\bm m_0) \rangle  
                    \langle u^{(0)}_{m\bm k}(\bm m_0) | \partial_{m_\alpha} u^{(0)}_{n\bm k}(\bm m_0) \rangle \notag \\
        &= J_{\rm ex}\sum_{m(\ne l)} 
           \frac{ \sigma^{l m}_{\alpha} \sigma^{m n}_{\beta}  }{ \varepsilon^{(0)}_{ml,\bm k}(\bm m_0) }
          -J_{\rm ex}\sum_{m(\ne n)} 
           \frac{ \sigma^{l m}_{\beta}  \sigma^{m n}_{\alpha} }{ \varepsilon^{(0)}_{nm,\bm k}(\bm m_0) }
\end{align}
Employing the above formulae, 
for instance, the derivative of the normalized Berry curvature $\Omega^{\bm k \bm m}_{n,i \beta}$ can read as 
\begin{align}\label{eq:omegaKM_diff}
    & \partial_{m_{\alpha}} {\overline \Omega}^{{\bm k}{\bm m}}_{n,i \beta} 
    = 2J_{\rm ex}  ~{\rm Im} \left[
                                     \sum_{l(\ne n)}
                                     \partial_{m_{\alpha}}
                                     \left(
                                     \frac{ v^{n l}_{i}  \sigma^{l n}_{\beta} }
                                          {( \varepsilon^{(0)}_{nl,\bm k}(\bm m_0) )^2 - \hbar^2\omega^{2} }
                                     \right)
                              \right] \notag \\
%                                     &= 2J    ~{\rm Im} \left[
%                                                         \sum_{l(\ne n)}
%                                                         \left(
%                                                               - \frac{\frac{\partial \omega^{2}_{nl} }{\partial m_{\alpha}}}
%                                                                      {(\omega^{2}_{nl} - \omega^{2})^2}
%                                                                      v^{n l}_{i}  \sigma^{l n}_{\beta}
%                                                               + \frac{ 
%                                                                        \frac{\partial v^{n l}_{i}}{\partial m_{\alpha}} \sigma^{l n}_{\beta}
%                                                                      + v^{n l}_{i} \frac{\partial \sigma^{l n}_{\beta}}{\partial m_{\alpha}}
%                                                                      }
%                                                                      {\omega^{2}_{nl} - \omega^{2}}
%                                                         \right)
%                                                         \right] \notag \\
% %                                                       
    &=-2J_{\rm ex}^{2}\sum_{l(\ne n)}{\rm Im} 
       \left[- 2
             \frac{\varepsilon^{(0)}_{nl,\bm k}(\bm m_0)  (\sigma^{n n}_{\alpha} - \sigma^{l l}_{\alpha})}
                  { (( \varepsilon^{(0)}_{nl,\bm k}(\bm m_0) )^2 - \hbar^2\omega^{2})^2 }
             v^{n l}_{i}  \sigma^{l n}_{\beta} \right. \notag \\
    &\quad \left. \ \ \ \ \ \ \ \ \ \ \ \ \ \ \ \ \ \ \ \ \
      + ~\sum_{m(\ne n)}      
         \frac{ \sigma^{n m}_{\alpha} v^{m l}_i \sigma^{l n}_{\beta} 
               + v^{n l}_i \sigma^{l m}_{\beta} \sigma^{m n}_{\alpha} }
              { \varepsilon^{(0)}_{nm,\bm k}(\bm m_0) 
               (( \varepsilon^{(0)}_{nl,\bm k}(\bm m_0) )^2 - \hbar^2\omega^{2}) } 
      - \sum_{m(\ne l)}
        \frac{  v^{n m}_{i} \sigma^{m l}_{\alpha} \sigma^{l n}_{\beta}
              + v^{n l}_{i} \sigma^{l m}_{\alpha} \sigma^{m n}_{\beta} }
             { \varepsilon^{(0)}_{ml,\bm k}(\bm m_0)
              (( \varepsilon^{(0)}_{nl,\bm k}(\bm m_0) )^2 - \hbar^2\omega^{2}) } 
       \right], 
\end{align}
and other derivative can be obtained as same as the above calculation.

\end{widetext}

\end{document}